\documentclass[a4paper,fleqn,usenatbib]{mnras}
\usepackage[T1]{fontenc}
\usepackage{ae,aecompl}
\usepackage{color,graphicx,amsmath,amssymb,natbib}
\usepackage[pagewise, mathlines]{lineno}
\bibpunct{(}{)}{;}{a}{}{,}
\newcommand{\ag}{Ann. Geophys.}

\newcommand{\asr}{Adv. Space Res.}
\newcommand{\vPUI}{v_\mathrm{PUI}}
\newcommand{\vPUIsc}{v_\mathrm{PUI}^{\mathrm{sc}}}
\newcommand{\vPUIsw}{v_\mathrm{PUI}^{\mathrm{sw}}}
\newcommand{\vrad}{v_\mathrm{r}}
\newcommand{\vSW}{v_\mathrm{sw}}
\newcommand{\wSC}{w_\mathrm{sc}}
\newcommand{\wSW}{w_\mathrm{sw}}

\newcommand{\Rd}{R_\mathrm{d}}

\title[ISN He, Ne, O density and PUI]{Solar cycle variation of interstellar neutral He, Ne, O density and pick-up ions along the Earth's orbit}
\author[J. M. Sok{\'o\l} et al.]{
Justyna M. Sok{\'o}{\l}$^{1}$\thanks{E-mail:
jsokol@cbk.waw.pl}, 
Maciej Bzowski$^{1}$, 
Marzena A. Kubiak$^{1}$, 
and Eberhard M{\"o}bius$^{2}$
\\
$^{1}$Space Research Centre, Polish Academy of Sciences, (CBK PAN), Warsaw, Bartycka 18A, Poland\\
$^{2}$University of New Hampshire, Durham, NH 03824, USA}

\date{Accepted 2016 March 01. Received 2016 February 27; in original form 2015 November 03.}

\pubyear{2016}

\begin{document}
\label{firstpage}
\pagerange{\pageref{firstpage}--\pageref{lastpage}}
\maketitle

%Abstract
\begin{abstract}
We simulated the modulation of the interstellar neutral (ISN) He, Ne, and O density and pick-up ion (PUI) production rate and count rate along the Earth's orbit over the solar cycle from 2002 to 2013 to verify if solar cycle-related effects may modify the inferred ecliptic longitude of the ISN inflow direction. We adopted the classical PUI model with isotropic distribution function and adiabatic cooling, modified by time- and heliolatitude-dependent ionization rates and non-zero injection speed of PUIs. We found that the ionization losses have a noticeable effect on the derivation of the ISN inflow longitude based on the Gaussian fit to the crescent and cone peak locations. We conclude that the non-zero radial velocity of the ISN flow and the energy range of the PUI distribution function that is accumulated are of importance for a precise reproduction of the PUI count rate along the Earth orbit. However, the temporal and latitudinal variations of the ionization in the heliosphere, and particularly their variation on the solar cycle time-scale, may significantly modify the shape of PUI cone and crescent and also their peak positions from year to year and thus bias by a few degrees the derived longitude of the ISN gas inflow direction. 
\end{abstract}


\section{Introduction}
After theoretical predictions in mid 1970-ties \citep{axford:72, vasyliunas_siscoe:76} and the first successful observations in mid 1980-ties by AMPTE \citep{mobius_etal:85a} and later by Ulysses \citep{gloeckler_etal:93a, geiss_etal:94a} pick-up ions (PUIs) have been used as a tool to investigate the environment inside and outside the heliosphere. The so-called interstellar PUIs \citep[e.g.][]{gloeckler_etal:04b, drews_etal:12a} are created through ionization of interstellar neutral (ISN) atoms by the solar wind and solar extreme ultraviolet (EUV) radiation. Together with direct sampling of ISN He (by Ulysses, \citealt{witte_etal:92a, witte:04} and IBEX, \citealt{mccomas_etal:09a, mobius_etal:09a}) and observations of the helium backscatter glow \citep[e.g.][]{weller_meier:74, vallerga_etal:04a} PUIs provide an important insight into the very local interstellar medium.

Before the IBEX launch, the inflow direction of the helium component of the ISN gas was determined to be $\lambda\sim255\degr$ (\citealt{witte:04}, ecliptic longitude, J2000 epoch; see also review by \citealt{mobius_etal:04a}). Analysis of the first two years of IBEX data \citep{bzowski_etal:12a, mobius_etal:12a} suggested that this direction may be different by $\sim4\degr$ towards larger values. This finding was supported by an analysis of PUI measurements from STEREO~A/PLASTIC \citep{drews_etal:12a}. The difference between the ISN inflow direction determined from the past and present measurements encouraged \citet{frisch_etal:13a, frisch_etal:15a} to review the determinations of the inflow direction of the ISN~He gas into the heliosphere during the last 40 yr. They gathered results obtained using different observational and analysis methods and pointed out that a change in the inflow longitude over time was statistically more likely than the lack of change. An important support of this hypothesis was the result of the analysis of the first two seasons of IBEX observations of ISN He, and the He, Ne, and O PUI count rate observed by STEREO. However, recent reanalysis of the Ulysses data \citep{bzowski_etal:14a, wood_etal:15a} and the newest analysis of IBEX data \citep{bzowski_etal:15a, leonard_etal:15a, mccomas_etal:15a, mccomas_etal:15b, schwadron_etal:15a} showed that the velocity vectors of the ISN gas derived from these two data sets are similar within uncertainties, with the temperature of the ISN gas higher than previously thought \citep[][]{witte:04}. Therefore the hypothesis of a temporal change in the ISN gas inflow direction is not supported. 

The results of the PUI analysis from the STEREO~A/PLASTIC measurements by \citet{drews_etal:12a} indicated an inflow ecliptic longitude different than the present consensus value of $\sim255\degr$. They were obtained from analysis of the He, Ne, and O~PUIs and the observed locations of the peak of the cone and crescent (an enhancement in the PUI count rate in the upwind direction, opposite to the peak of the downwind focusing cone). In this paper, we look into details of the production of He, Ne, O PUIs to test if the apparent shifts in the ISN longitude observed by \citet{drews_etal:12a} may be related to variations in the ionization of interstellar gas due to solar activity changes. The differences in the inflow direction derived from ISN gas and PUI measurements motivate us to study how accurate the determination of the neutral interstellar flow direction is when it is based on the position of the downwind cone and upwind crescent of the PUIs close to the Sun observed by a spacecraft orbiting the Sun similarly as Earth does. The key question is whether effects of time- and latitude-variation of the ionization rate on the ISN gas distribution along the Earth's orbit can masquerade as a change of the inflow direction. 

We performed a theoretical study of the structure of the ISN He, Ne, and O gas density and their PUIs at the Earth's orbit\footnote{By `at Earth', `at the Earth's orbit' etc. we mean the locations precisely at the instantaneous distance of the Earth from the Sun, not 1~au. The reason for this differentiation will be explained later in the text.} and of the evolution of this structure during solar cycle. We assumed a constant inflow direction of ISN gas in front of the heliosphere and calculated the density of the gas inside Earth's orbit from 2002 to 2013. Based on these densities, the instantaneous local production rate of PUIs  at Earth was determined, and subsequently the expected count rates of PUIs measured by an electrostatic analyzer were calculated. We studied the factors responsible for modulation of the modelled quantities with special attention to the modification of the observed signatures by the ionization rates close to the Sun and the presence of the crescent at the upwind side. Finally, we determined the longitude of the peak locations approximating the cone and crescent peaks with a Gaussian function and compared them with the longitude of the ISN flow assumed in the simulation of the ISN gas density.

Section~\ref{sec:Models} presents details of the model. In Section~\ref{sec:Densities}, the densities of ISN He, Ne, and O are shown. Section~\ref{sec:PUIs} presents the PUI productions rates at Earth and the expected count rates. The peak locations from the Gaussian fit are shown in Section~\ref{sec:fits}. Section~\ref{sec:summary} summarizes the results and concludes the study. 

\section{Models and simulations}
\label{sec:Models}
\citet{rucinski_etal:03} presented an analysis of the expected modulation of the ISN He density and PUI flux in the heliosphere as a function of distance from the Sun over a solar activity cycle. We extend their analysis to Ne and O, and in contrast to \citet{rucinski_etal:03}, who took into account only spherically symmetric photoionization losses, we include all the most relevant ionization processes, it is photoionization by solar EUV radiation, charge exchange with solar wind particles, and electron-impact ionization. The total ionization rate is a sum of photoionization, which is mostly responsible for the global, solar-cycle related modulation of the total ionization rate, electron-impact ionization, which is responsible for slight departures of the total ionization rate from the inverse-square dependence on the heliocentric distance and contributes to the heliolatitudinal structure in the ionization field, and charge exchange between ISN atoms and solar wind ions, which is mostly responsible for the heliolatitude dependence of the total ionization rate. All three reactions feature significant daily and monthly quasi-random fluctuations. We include the deviation of the ionization rate from spherical symmetry due to the latitudinal dependence of the solar wind speed and density (see \citealt{sokol_etal:13a}, and for a most recent review \citealt{sokol_etal:15d}). In our analysis we use a model of ionization rates based on observations \citep{bzowski_etal:13b, sokol_etal:13a, bochsler_etal:14a, sokol_bzowski:14a}. Time-dependent effects in the distribution of ISN gas were studied by \citet{rucinski_bzowski:95b, rucinski_bzowski:96}, \citet{bzowski_etal:97}, and \citet{bzowski:03} for H and by \citet{rucinski_etal:03} for He, but to our knowledge, the time evolution of the ISN Ne and O density has not been available in the literature. As we will show, even seemingly small effects of solar cycle modulated ionization are important, especially for ISN~O.

The densities of ISN He, Ne, and O were calculated using the Warsaw Test Particle Model (see \citealt{bzowski_etal:12a, kubiak_etal:14a}, and for recent review \citealt{sokol_etal:15b}) for a Maxwell--Boltzmann distribution of the ISN gas in front of the heliosphere. This model follows the original `hot model' paradigm \citep{fahr:78, thomas:78, wu_judge:79a}, modified to account for ionization losses dependent on time and heliolatitude. In this paper, we focused on one full solar activity cycle (SC) from the maximum of SC~23 ($\approx2002$) to the maximum of SC~24 ($\approx2013$). We chose this time interval to have a homogeneous time series of daily photoionization rates, which are available as a continuous and long time series only from TIMED/SEE \citep{woods_etal:05a}, launched in 2002. The photoionization rates were calculated by direct integration of the solar EUV flux measurements from TIMED/SEE (Level3 data, version 11) multiplied by the cross-section from \citet{verner_etal:96}; for a detailed discussion of the photoionization models see \citet{bochsler_etal:14a} and \citet{sokol_bzowski:14a}. For photoionization rates, the pole-to-equator flattening of the latitudinal structure was adopted after \citet[][see also discussion in \citealt{bzowski_etal:13a}]{auchere_etal:05a, auchere_etal:05b}. The latitudinal structure of the solar wind proton speed and density was adopted from \citet{sokol_etal:13a}, with the OMNI \citep{king_papitashvili:05} solar wind time series for the ecliptic plane. We assumed a $1/r^2$ decrease with distance from the Sun $r$ for the solar wind density and photoionization rates and we assumed that solar wind speed does not change over the distances from the Sun that we studied. The electron impact ionization rate was calculated following a model proposed by \citet{rucinski_fahr:89, rucinski_fahr:91}, adapted originally for ISN~H by \citet{bzowski:08a}, and developed for ISN~He, Ne, and O by \citet{bzowski_etal:13b}. In this model, a measurement-based relation of the decrease of the electron impact ionization rate with distance from the Sun was adopted, which does not follow the $1/r^2$ relation (see fig.~2 in \citealt{bzowski_etal:13b}). In effect, this source of ionization becomes important mostly inside $\sim 2$~au from the Sun. In the calculation of the ISN gas density in the heliosphere we used ionization rates averaged over solar (Carrington) rotation\footnote{1 sidereal solar rotation = 27.2753~d (Carrington rotation, CR afterward)} periods linearly interpolated to daily values.

\begin{figure*}
	\begin{tabular}{cc}
	\includegraphics[scale=0.6]{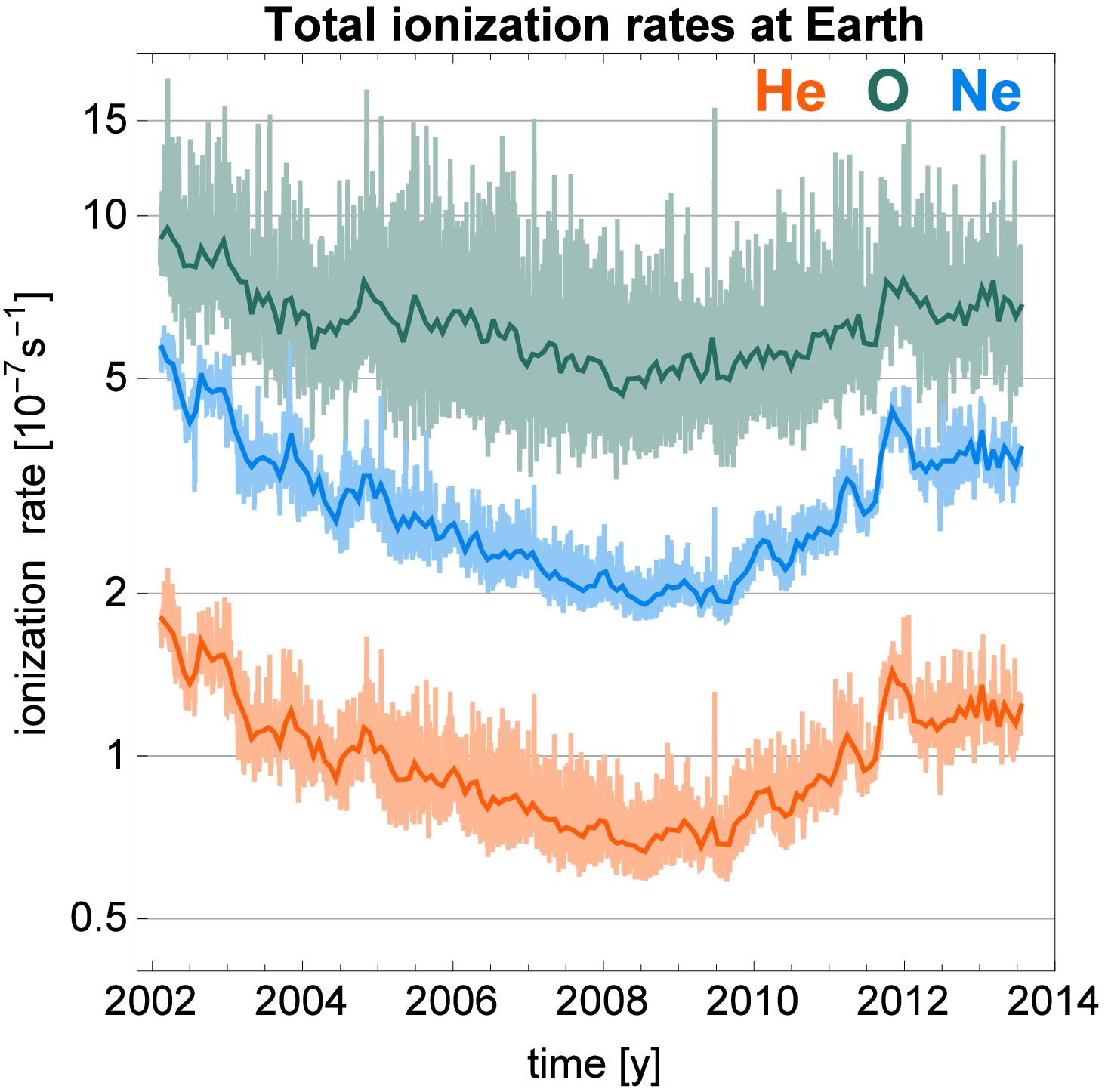}
	\includegraphics[scale=0.585]{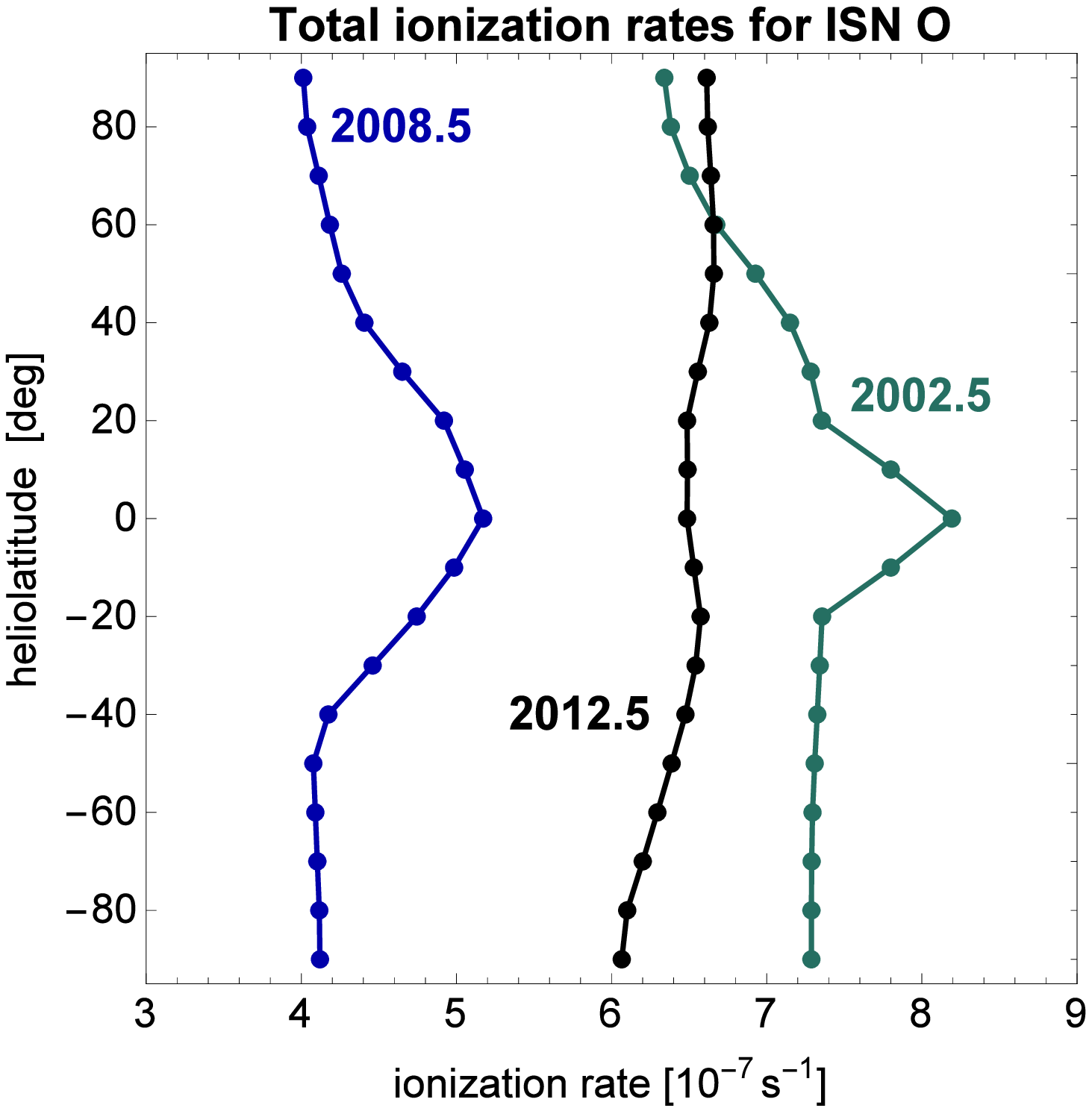}
	\end{tabular}
	\caption{Left-hand panel: time series of daily (faint and strongly varying lines) and CR-averaged (dark lines) total ionization rates for He, Ne, and O at Earth. Right-hand panel: latitudinal structure of the total ionization rates for O at 1~au close to the maximum of SC~23 (2002.5) and SC~24 (2012.5), and for the minimum of SC~23/24 (2008.5) (for this figure only, the stationary atom approximation is used).}
	\label{figIonRatesBetaTot}
\end{figure*}

The latitudinal structure of the ionization rates affects mostly the charge exchange ionization process (due to the latitudinal variation of the solar wind, right-hand panel of Fig.~\ref{figIonRatesBetaTot}), which is an significant source of ionization losses for ISN~O, contributing up to $40\%$ of the total ionization rate in the ecliptic plane. For ISN~Ne, charge exchange is less important, and for ISN~He almost negligible (see fig.~1 in \citealt{bzowski_etal:13b}). The left-hand panel in Fig.~\ref{figIonRatesBetaTot} presents the total ionization rate for He, Ne, and O at the Earth's orbit as a daily and CR-averaged time series. The figure presents time variations over the interval studied in this analysis. The daily series varies rapidly due to the short-scale fluctuations of the solar wind and EUV flux, and the general trend is illustrated by the CR averages. The rates for He and Ne show clear variations with the solar activity cycle, while for O these variations are less pronounced. This is due to the large contribution of charge exchange with solar wind protons to the total ionization rate for O, which follows the evolution of the solar wind flux with time. The solar wind flux featured a secular drop during the previous two solar activity cycles \citep[e.g.][]{mccomas_etal:08a, mccomas_etal:13b, sokol_etal:13a}, and close to the ecliptic plane it does not vary quasi-periodically with solar activity. Consequently, the quasi-periodic component in the ionization rate for O is weaker than in the ionization rates for He and Ne, which have a dominant contribution from photoionization rate, strongly modulated by solar activity (see also discussion in \citealt{bzowski_etal:13b}). The total ionization rate increases from a minimum to maximum value by a factor of about $2.5$ for He, $2.8$ for Ne, and $2$ for O. Even though the amplitude of variation of the ionization rate is the weakest for O, the total rate is the highest, which results in the largest attenuation of the ISN~O gas measured in the Earth's orbit.
% ionization rates

The right-hand panel in Fig.~\ref{figIonRatesBetaTot} shows the differences in the latitudinal structure of the total ionization rates at 1~au for ISN~O for three selected years (in this case only, in the approximation of stationary atoms\footnote{In the stationary atom approximation, the charge exchange collision speed is adopted as equal to the solar wind expansion speed. In reality, a more accurate approximation would use the magnitude of the vector difference between the radial speed of solar wind and the instantaneous orbital velocity of the atom.}). During solar minimum, the solar wind exhibits a bi-modal structure, with a slow and dense streams at low latitudes, and a fast and dilute flow at high latitudes, with a transition region at mid-latitudes. During a short interval of a few months at solar maximum, the slow solar wind extends to mid- and high latitudes and the structure of the solar wind outflow is almost uniform in latitude. The modulation in the solar wind structure in the northern and southern hemispheres is shifted in phase (see fig.~20 in \citet{sokol_etal:13a}, figs~7 and 11 in \citet{sokol_etal:15d}, and \citet{tokumaru_etal:15a}). In effect, the solar wind structure is approximately spherically symmetric only for a very short time during the solar cycle, and, as illustrated in the right-hand panel of Fig.~\ref{figIonRatesBetaTot}, the charge exchange ionization rate is uniform in heliolatitude also during solar maximum. Many of the ISN atoms detected in the ecliptic plane traverse also higher latitudes during their travel from the front region of the heliosphere. Therefore the latitudinal structure of the ionization rate, especially for ISN~O, has to be appropriately taken into account in the calculation of the ISN gas distribution, especially on the downwind side.

Our objective is to test how the PUI count rate, measured by a spaceborne detector in the Earth's orbit, reflects the ISN gas flow direction in front of the heliosphere. We used the ISN gas parameters derived from the analysis of IBEX measurements of ISN~He in 2009 and 2010 after \citet[][longitude $259.2\degr$, latitude $5.12\degr$ - ecliptic coordinate system for J2000 epoch - speed $22.756$~km~s$^{-1}$, and temperature 6165~K]{bzowski_etal:12a}, with the longitude similar to that found by \citet{drews_etal:12a} from the analysis of PUI measurements. We used these values in all simulations, unless stated otherwise. These ISN parameters are now updated in the analysis of \citet{mccomas_etal:15b}, \citet{bzowski_etal:15a}, and \citet{schwadron_etal:15a}, but in our analysis we decided to keep the direction consistent with the result of the PUI study presented by \citet{drews_etal:12a} to test whether this direction might be shifted in longitude by ionization processes. We started our test particle calculations with the gas flowing from the source region (set at 150~au from the Sun) and searched how the bulk flow is modified between the source region in front of the heliosphere and the Earth's orbit by the gravitational focusing and the ionization field in the heliosphere. We worked in the Sun-centred ecliptic coordinate frame and we maintained this convention throughout the paper. We carried out the analysis based on one fixed, arbitrary set of the inflow parameters, but we checked that the conclusions do not depend on the adopted ISN flow direction.

\section{Densities of ISN He, Ne, and O}
\label{sec:Densities}
\begin{figure*}
	\begin{center}
	\includegraphics[scale=0.65]{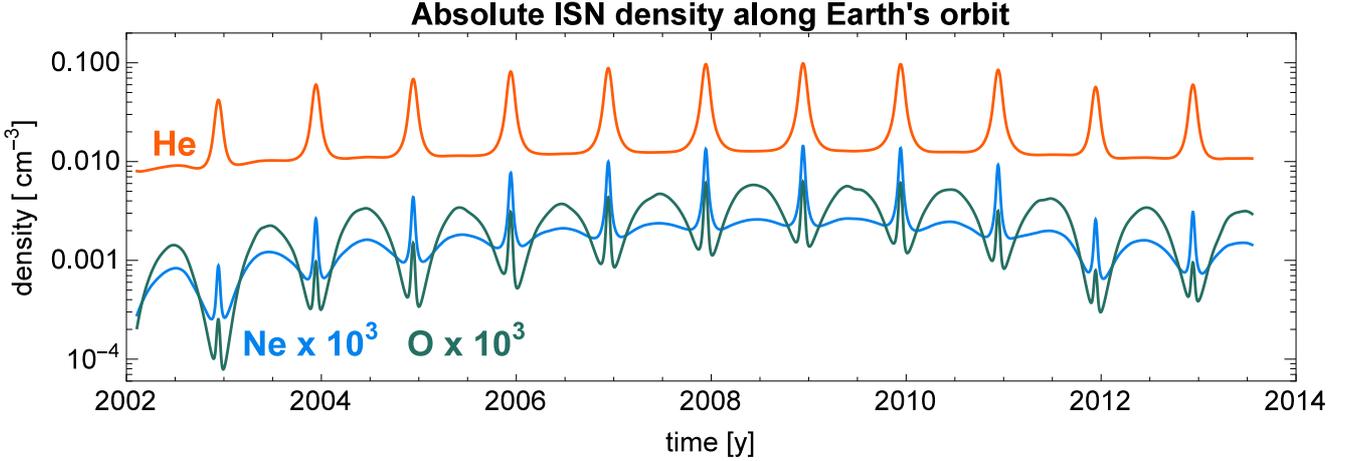}
	\end{center}
	\caption{Time series of absolute ISN He, Ne, and O density calculated along the Earth's orbit. The ISN Ne and O densities are scaled upward by a factor of $10^3$. The vertical scale is logarithmic. The solar cycle related systematic effects for He can be better appreciated in Fig.~\ref{figHeInflowSpeedDens}.}
	\label{figDensSeriesHeNeOx}
\end{figure*}

\begin{figure}
	\begin{center}
	\includegraphics[width=\columnwidth]{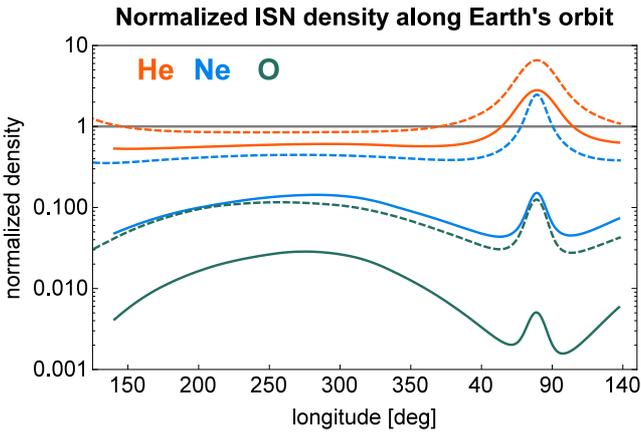}
	\end{center}
	\caption{Density of ISN He, Ne, and O calculated along the Earth's orbit for each day of 2002 (solid lines) and 2008 (dashed lines), normalized to the respective densities of these species in the LIC. The horizontal line emphasizes the differences in the densities relative to the LIC.}
	\label{figDensBase}
\end{figure}

In the first step, we calculated the densities of the ISN species assuming the following densities in the source region in the Local Interstellar Cloud (LIC): $0.015$~cm$^{-3}$ for He,  $0.582 \times 10^{-5}$~cm$^{-3}$ for Ne, and $0.5 \times 10^{-4}$~cm$^{-3}$ for O \citep{slavin_frisch:07a, slavin_frisch:08a}. These densities result from the abundance of these species in interstellar matter on one hand and from the ionization state of these species in the LIC on the other hand. The densities inside the heliosphere were simulated at a one-day resolution from 2002 February 9 through 2013 July 20, but with a CR average applied to the ionization rates (dark lines in the left-hand panel of Fig.~\ref{figIonRatesBetaTot}). Fig.~\ref{figDensSeriesHeNeOx} shows the full sequence of the resulting density of ISN He, Ne, and O observed along the Earth's orbit as a function of time. In Fig.~\ref{figDensBase}, we compare normalized densities for the solar minimum ($\sim 2008$) and maximum ($\sim 2002$) phase, shown as a function of the Earth's ecliptic longitude. In this figure one can study in detail differences between the densities of the three species in the cone region due to different thermal speeds and different ionization losses. The simulated densities show a clear downwind focusing cone for all three species and both activity phases. The upwind crescent is well visible for O during both solar minimum and maximum, for Ne also, but during solar minimum it is very faint, and for He the crescent is absent for both high and low activity conditions. The ISN Ne and O densities  are similar and less abundant at Earth than He. Fig.~\ref{figDensBase} shows the scale of attenuation of these species with respect to the values in the LIC. ISN O is the most attenuated, ISN Ne is also strongly reduced except in the cone region during solar minimum, when the density enhancement due to the focusing by solar gravity exceeds the ionization losses. The increase of the ISN He density is the strongest, almost $\sim6.5$ times in the cone during solar minimum. The difference in the absolute densities between solar minimum and maximum ranges from $\sim 29\%$ in the upwind region to $\sim 58\%$ in the downwind region for He, $\sim 68\%$ and $\sim 94\%$ for Ne, and for O $\sim 75\%$ and $\sim 96\% $, respectively.
% absolute denisty time series
% density pattern

Fig.~\ref{figDensSeriesHeNeOx} illustrates solar cycle related trends in the density distribution along the Earth's orbit that are due to the solar cycle related changes in the absolute ionization rates: the densities increase when the ionization rates decrease and opposite, the densities decrease when the ionization rates increase (see also Fig.~\ref{figIonRatesBetaTot}). The minimum in the ionization rate occurred about 2008 and shortly afterwards the general upward trend in the densities reversed. These effects are observed for all species, but they are the strongest for O, and next for Ne. As a result, the densities are not symmetrical around the crescent peaks when observed by an observer moving with the Earth.

\begin{figure*}
		\begin{center}
		\begin{tabular}{c}
		\includegraphics[scale=0.65]{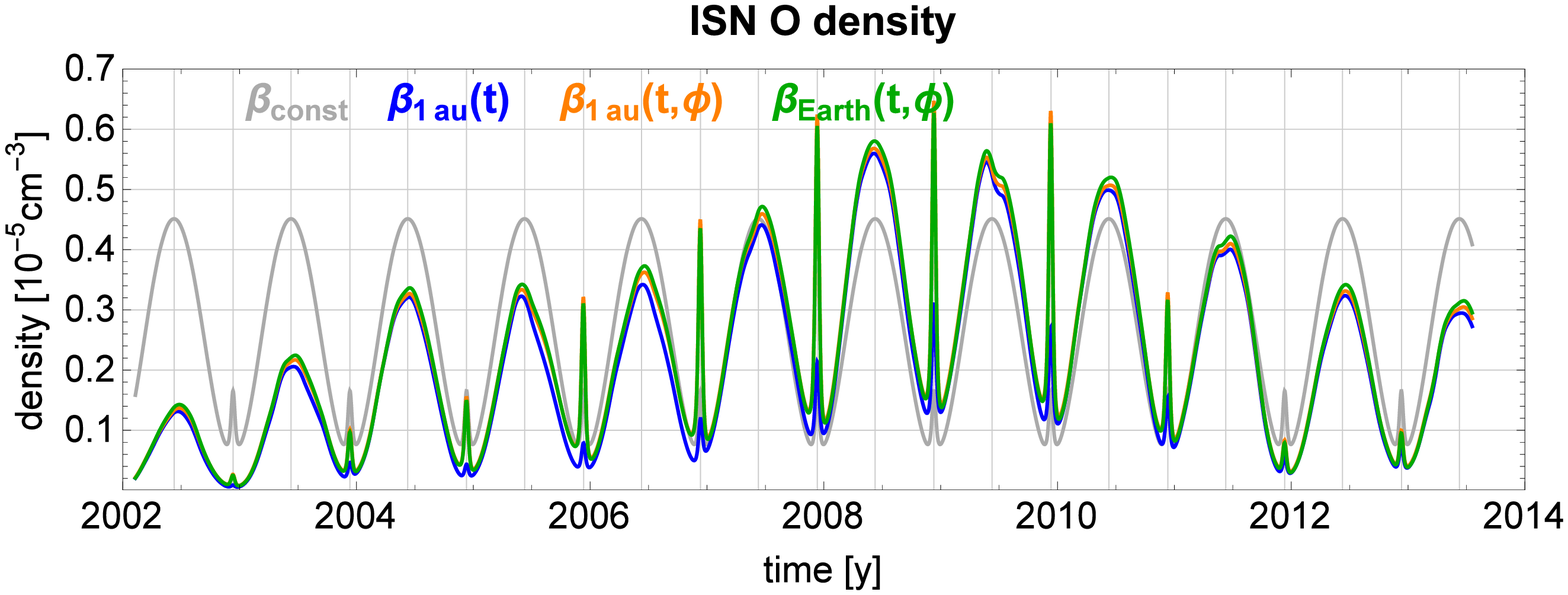}\\
		\includegraphics[scale=0.65]{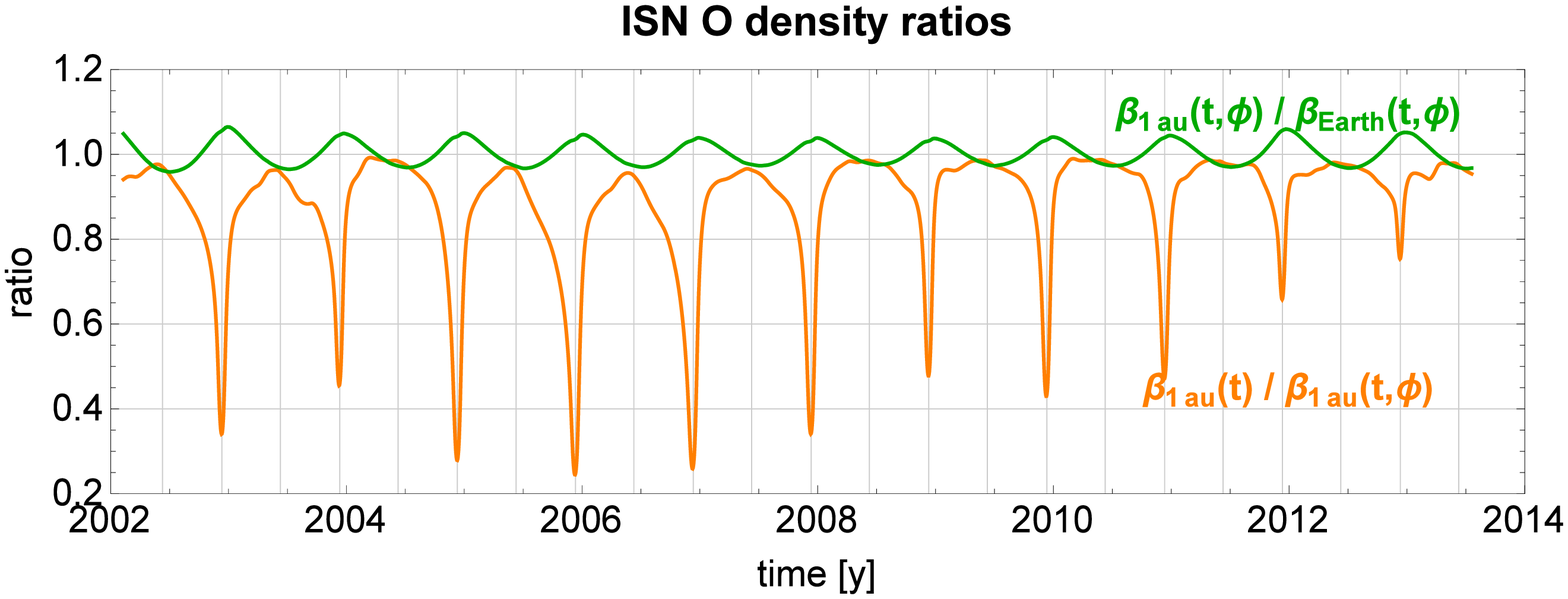}\\
		\end{tabular}
		\end{center}
		\caption{Top panel: modification of the ISN~O density calculated under different assumptions and simplifications. $\beta_{\mathrm{const}}$ -- calculations with constant, spherically symmetric in-ecliptic ionization at 1~au (grey); $\beta_{\textrm{1au}} \left( t \right)$ -- spherically symmetric, time variable in-ecliptic ionization rates at 1~au for the time of detection (blue); $\beta_{\textrm{1au}} \left( t,\phi \right)$ -- ionization variable in time and heliolatitude at 1~au (orange); $\beta_{\mathrm{Earth}} \left( t, \phi \right)$ -- same as before, but for the true Earth's distance to the Sun (green). Bottom panel: ratios of the results for spherically symmetric versus latitudinally dependent ionization rates (orange line) and for a circular 1~au orbit versus the realistic Earth's orbit (green line). The vertical grid lines mark the times of the crossing of the longitudes of upwind and downwind directions.} 
	\label{figDensityOxEarthFull}
\end{figure*}

Next, we studied the effect of some assumptions and simplifications commonly made in the modelling of the density pattern, taking ISN~O as an example. The ionization losses are one of the modelling aspects that is often simplified as constant in time and spherically symmetric. Also the Earth's orbit is typically assumed to be circular, instead of elliptical. We start with the simplest assumption of a circular Earth's orbit (with the radius equal to 1~au), and a constant, spherically symmetric ionization rate of $\sim 2.8 \times 10^{-7}$ (i.e., the average value of the in-ecliptic photoionization rates from 2005.5 to 2008.5; $\beta_{\mathrm{const}}$). The resulting ISN~O density is illustrated with the grey line in the top panel of Fig.~\ref{figDensityOxEarthFull}. Due to the absence of time variations in the ionization rate, the density time series is strictly periodic and the only visible features are the very narrow focusing cone downwind and the broad crescent upwind. In the next step, we introduced the time variations of the in-ecliptic ionization rates while still keeping the uniform distribution with heliolatitude and the circularity of the Earth's orbit ($\beta_{\mathrm{1au}}(t)$, blue line in the figure). Now the magnitudes of the upwind and downwind density peaks are modulated in time and follow the solar cycle, with higher densities during solar minimum, and lower during solar maximum. Additionally, the cone and the crescent are no longer symmetric with respect to the respective peak values due to time variations of ionization rates.
% density for oxygen

The orange line in Fig.~\ref{figDensityOxEarthFull} represents the density of ISN~O calculated with the heliolatitude $\phi$ dependence of the ionization rates included ($\beta_{\mathrm{1au}}(t,\phi)$) in addition to the time dependence. The densities are calculated for the in-ecliptic orbit, however, the total density is accumulated from a broad source region in the LIC, focused by solar gravitation, and thus a part of the population has traversed also higher latitudes. Therefore, the latitudinal structure of the ionization inside the heliosphere, which is controlled mainly by the solar cycle dependent variations of the solar wind speed and density with heliolatitude, is important in the modelling especially for ISN~O, for which the charge exchange losses have a significant contribution to the total ionization rates (see e.g. fig.~1 in \citealt{bzowski_etal:13b}). The orange line in the bottom panel of Fig.~\ref{figDensityOxEarthFull} presents the ratio of the ISN~O density when comparing the cases with the spherically symmetric ionization with the heliolatitude-dependent ionization. The ratio differs from 1 by a few percent in the crescent peak, but decreases rapidly in the crescent slopes and reaches a minimum in the cone. This illustrates the high importance of the latitudinal structure of the solar wind for the density of ISN~O in the cone region.

The effect of the last commonly made simplification, i.e., adopting a circular 1~au orbit instead of the realistic elliptical Earth's orbit ($\beta_{\mathrm{Earth}}(t,\phi)$) is illustrated by green lines in Fig.~\ref{figDensityOxEarthFull}. The eccentricity modifies the calculated densities slightly, introducing an annual modulation by up to $7\%$. The eccentricity modifies the densities for given portions of the time series significantly enough to be taken into account in a detailed analysis of the ISN densities, but its influence is much less than that of the other effects.

The structure of the crescent is sometimes corrugated in the peak (see e.g., the crescent in the middle of 2007, 2009, and 2011 of top panel of Fig.~\ref{figDensityOxEarthFull}). This wavy structure is due to short-lasting temporal changes in the ionization rates that occurred in the periods when the Earth traverses the crescent region of its orbit. The crescent is broad, it covers far more than $180\degr$ in longitude along the orbit, and traversing it takes the Earth more than 6 months, so a few percent increase/decrease in the ionization rates during $\sim2$ months can make the shape wavy. Additionally, asymmetries in the density time series appear around the cone (compare the two minima around the cone bottom in the blue line in the top panel of Fig.~\ref{figDensityOxEarthFull} in, e.g., 2007 or 2010). These are also caused by short-time variations in ionization. The cone traversal by the Earth lasts slightly longer than a month, and the month-to-month differences in the ionization rate can significantly modify the slopes of the narrow cone structure. 

While the cone is always present for the species under discussion, it may be missed by an observer in the Earth's orbit because the ISN density is strongly reduced by ionization for heavy species, which results in a faint signal compared to magnitudes before and after the cone (see e.g. 2012 in top panel of Fig.~\ref{figDensityOxEarthFull}). Note also that due to the inclination of the inflow direction to the ecliptic by $\sim 5\degr$, the Earth misses the absolute maximum of the density pattern in the cone.

\begin{figure}
	\begin{center}
	\includegraphics[width=\columnwidth]{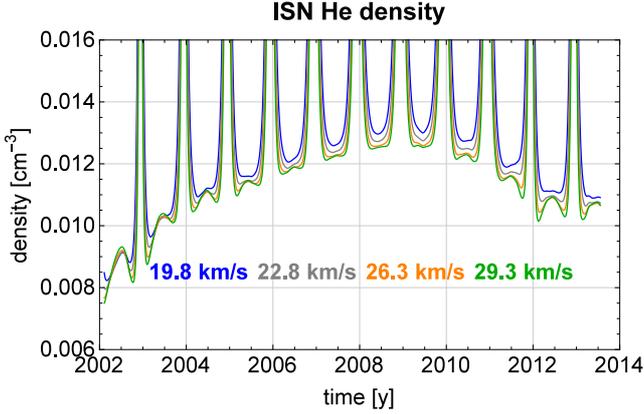}
	\end{center}
	\caption{Density of ISN~He calculated in the Earth's orbit for various inflow speeds of ISN gas in front of the heliosphere. A crescent builds up during higher solar activity and for faster speeds.}
	\label{figHeInflowSpeedDens}
\end{figure}

The crescent is not present in the density series for ISN~He in Earth's orbit around the Sun even for the solar minimum years, contrary to the density of ISN~Ne and O (Figs~\ref{figDensSeriesHeNeOx}, \ref{figDensBase}, and \ref{figHeInflowSpeedDens}). This motivated us to study the effect of the ISN bulk speed on the creation of the crescent in the ISN He density. In Fig.~\ref{figHeInflowSpeedDens} we compare time series of the ISN~He density in the Earth's orbit for four selected speeds of the ISN gas in front of the heliosphere, while keeping the remaining parameters unchanged. For higher ISN flow speed, the crescent is stronger, which suggests that the formation of the crescent for ISN~He is somewhat sensitive to the Mach number of the flow of a given species (the higher Mach number, the more collimated is the ISN beam), especially for the years with strong ionization losses, when the small speed portion of the population is more readily ionized. We also verified that higher ionization leads to a more structured crescent. 
% density for He with various inflow speeds

In addition, we studied whether the distance of observation from the Sun has an effect on the presence of the crescent in the ISN~He density in the ecliptic plane. We calculated the ISN~He density for a number of distances down to $0.4\Rd$, where $\Rd$ is the actual Earth's distance from the Sun and we concluded that the formation of the crescent is weakly distance-dependant. For ISN~He in the ecliptic plane, the density crescent appears for the closest distance from the Sun and only during the phase of high solar activity.

\section{Signatures of pick-up ions as seen from the Earth's orbit}
\label{sec:PUIs}
Direct measurements of the absolute density of the ISN gas close to the Sun are not carried out, but in-direct methods have been successfully applied to derive the density of ISN gas in the heliosphere, e.g.; analysis of the ISN~He flux measured by Ulysses/GAS \citep{witte_etal:96, witte:04} or analysis of the He$^{2+}$ PUI fluxes from Ulysses/SWICS \citep{gloeckler_etal:04b}. Direct sampling experiments, like Ulysses or IBEX, measure the differential flux as a function of the location in the sky while integrating over the energy, and the measurements can only be successfully carried out on fragments of the spacecraft orbit around the Sun \citep{mobius_etal:12a}. Observations of the helioglow, both spectroscopic and photometric, measure a convolution of the ISN density and the illuminating solar EUV flux along the line-of-sight (LOS), modulated by the radial component of the velocity vector of the gas. The extraction of the absolute density is biased by the integration along the LOS. Thus, observations of derivative populations, like PUIs, are very useful to study the ISN gas \emph{in situ}.

In this section, we present simulations of the expected He, Ne, and O PUIs along the Earth's orbit. For the calculations we used the daily time series of the ISN gas densities (Fig.~\ref{figDensSeriesHeNeOx}) and ionization rates (Fig.~\ref{figIonRatesBetaTot}) in the ecliptic plane, between $0.4\Rd$ and $1\Rd$ in $0.1\Rd$ increments. To calculate the distance dependence of the ionization rates we assumed solar wind speed constant with heliocentric distance and an inverse-square law with distance for solar wind density and EUV flux \footnote{More in Appendix~\ref{sec:appendixPart2}.}. We approximated the PUI observations for a detector at rest (it is a detector that has a relative velocity to the Sun equal zero) and has an exactly sunward, pencil-beam like (infinitesimally small and radial) field of view (FoV). Determination of the PUI formulae used in the calculations are presented in Appendix~\ref{sec:appendix}, and details of the calculation scheme are presented in Appendix~\ref{sec:appendixPart2}.

In the first step, we calculated the instantaneous, local PUI production rate (the source function, e.g., \citealt{vasyliunas_siscoe:76, mobius_etal:88a, mobius_etal:95, bzowski_rucinski:96a}) at Earth as a function of distance $r$ from the Sun and time $t$:
	\begin{equation}
	S(r,t)=n(r,t)\beta(r,t),
	\label{eq:ProdRate}
	\end{equation}
where $n(r,t)$ is the density of the ISN atoms and $\beta(r,t)$ the total ionization rate at time $t$ and solar distance $r$. The temporal and spatial variation along the orbit of the virtual detector are used interchangeably in the further discussion because the longitude grid is based on a daily grid (the Earth's location changes by about $1\degr$ in ecliptic longitude each day, although not exactly linearly). The calculations were based on a daily grid identical to that used for the calculations of density discussed in the previous section.
% zbiorczy rysunek prodRates, totalFluxR, partialFluxwSC
	\begin{figure*}
	\centering	
	\begin{tabular}{ccc}
	\includegraphics[scale=0.3]{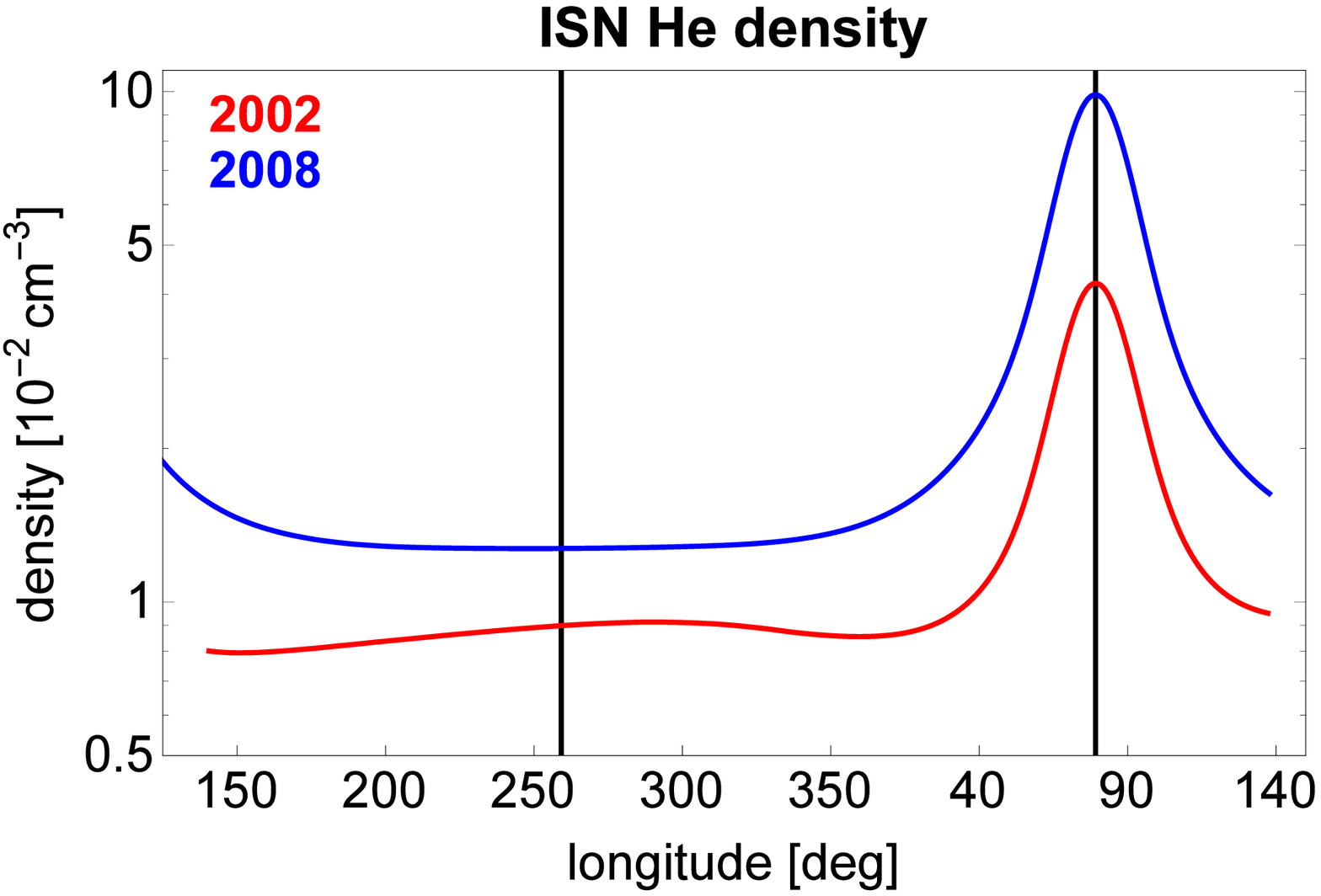} &	\includegraphics[scale=0.3]{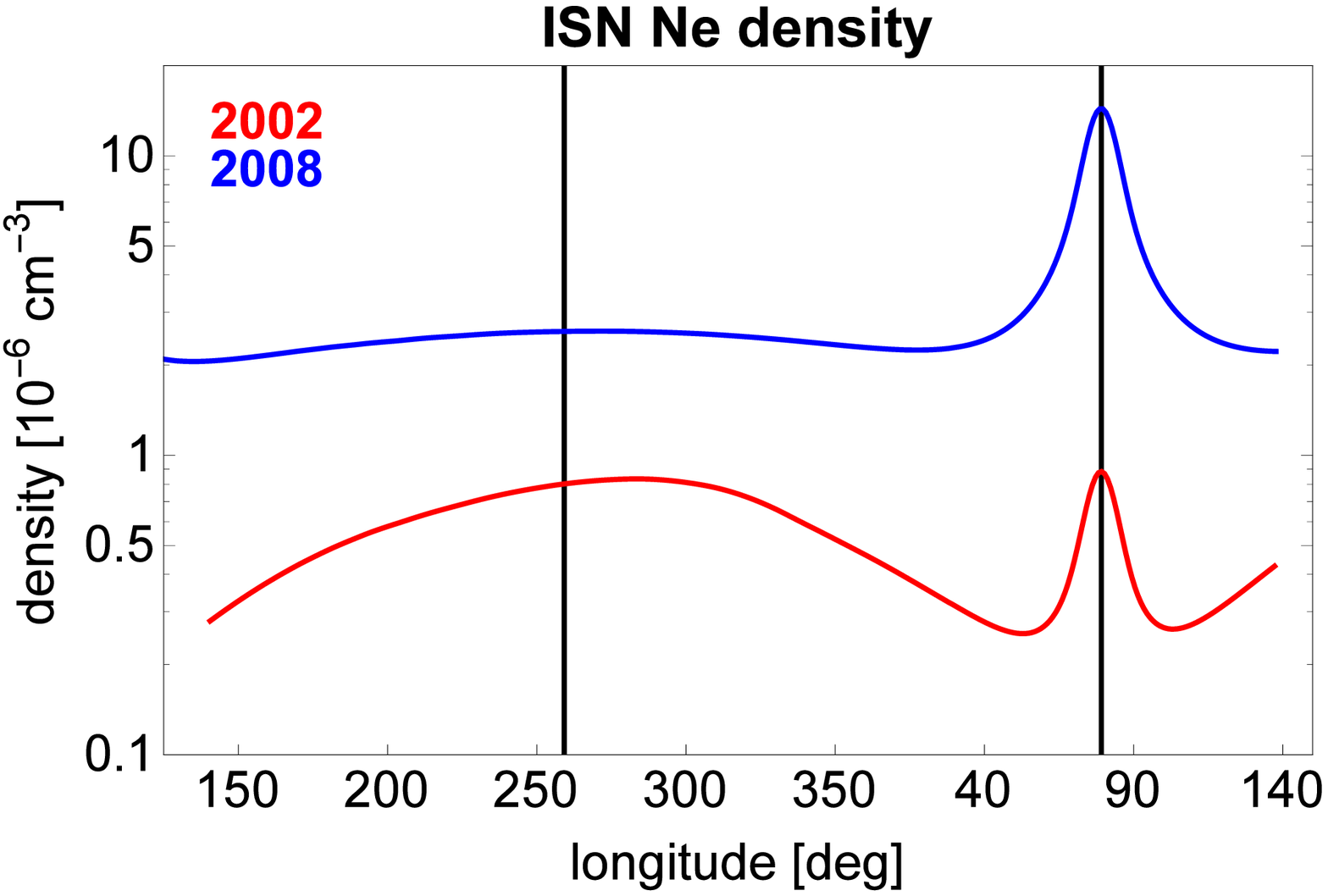} &
	\includegraphics[scale=0.305]{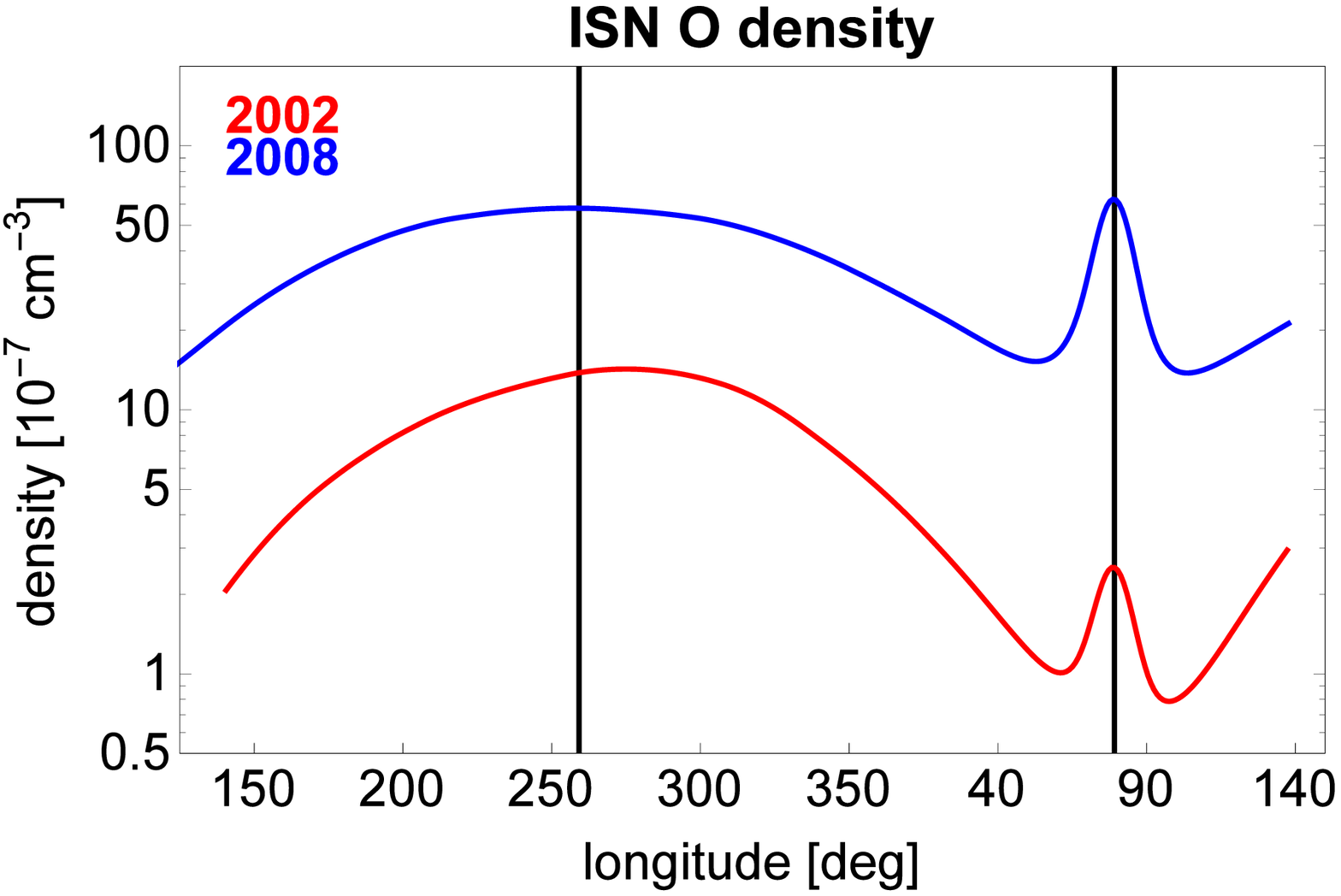} \\
	\includegraphics[scale=0.3]{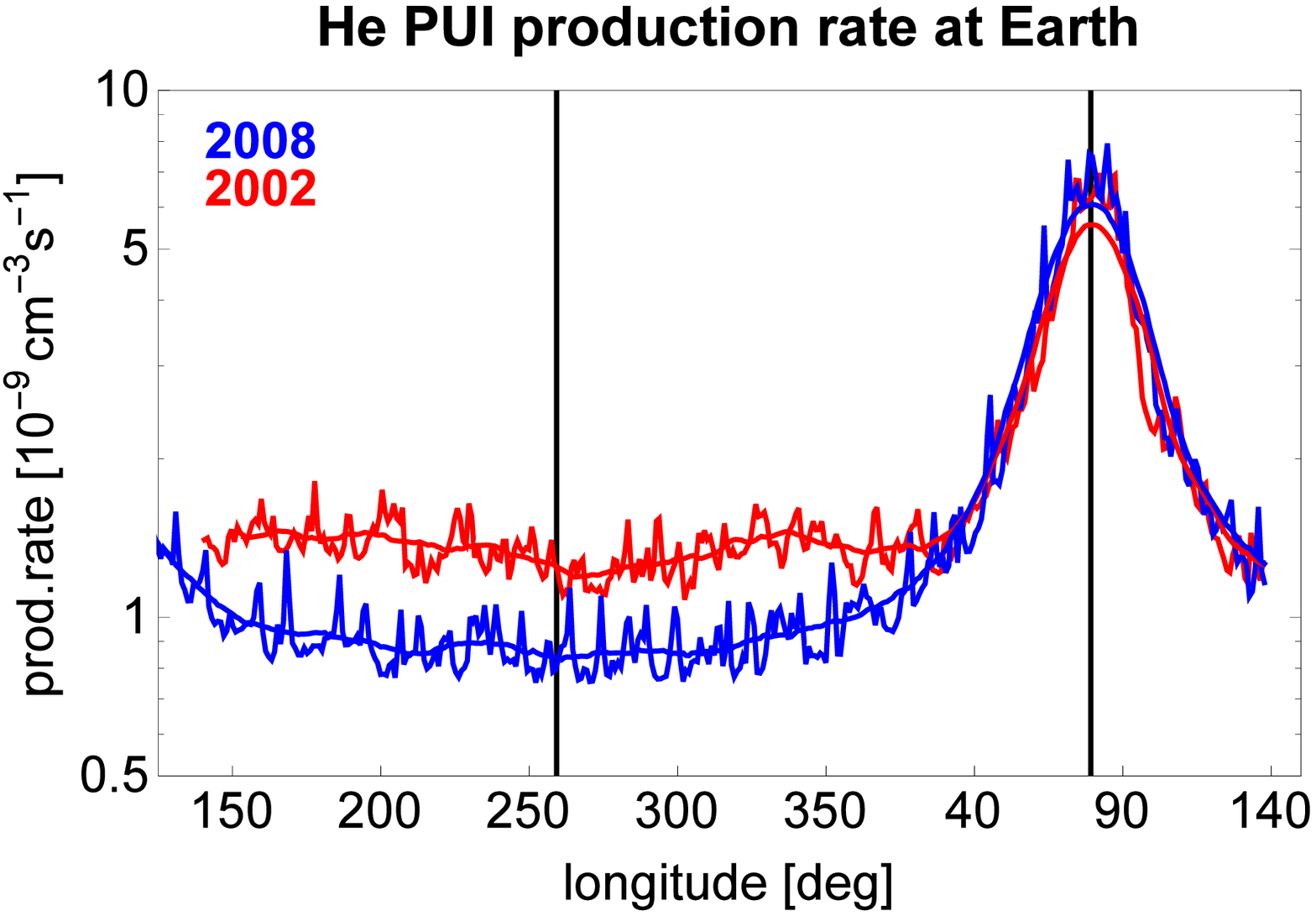} &	\includegraphics[scale=0.3]{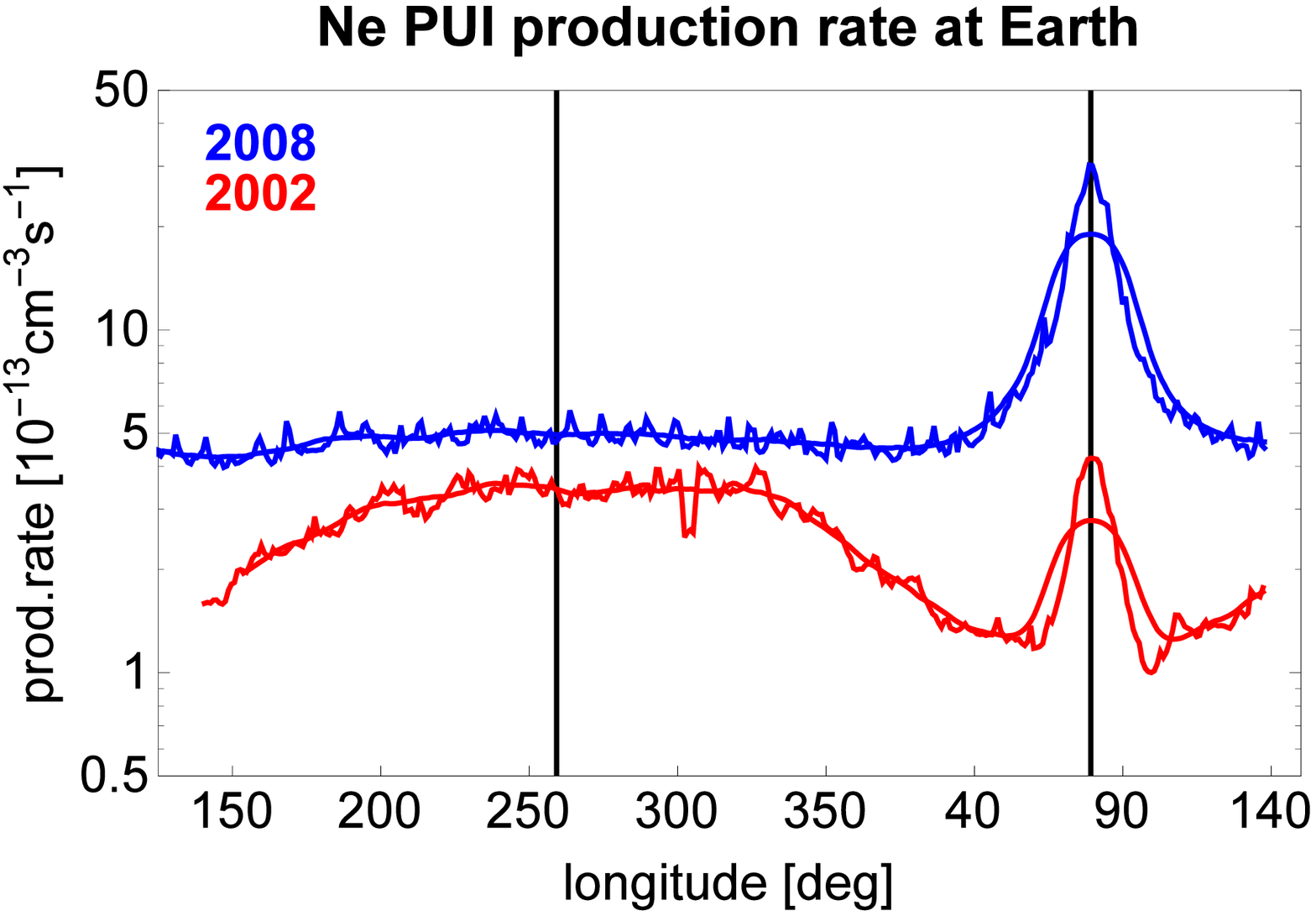} &
	\includegraphics[scale=0.3]{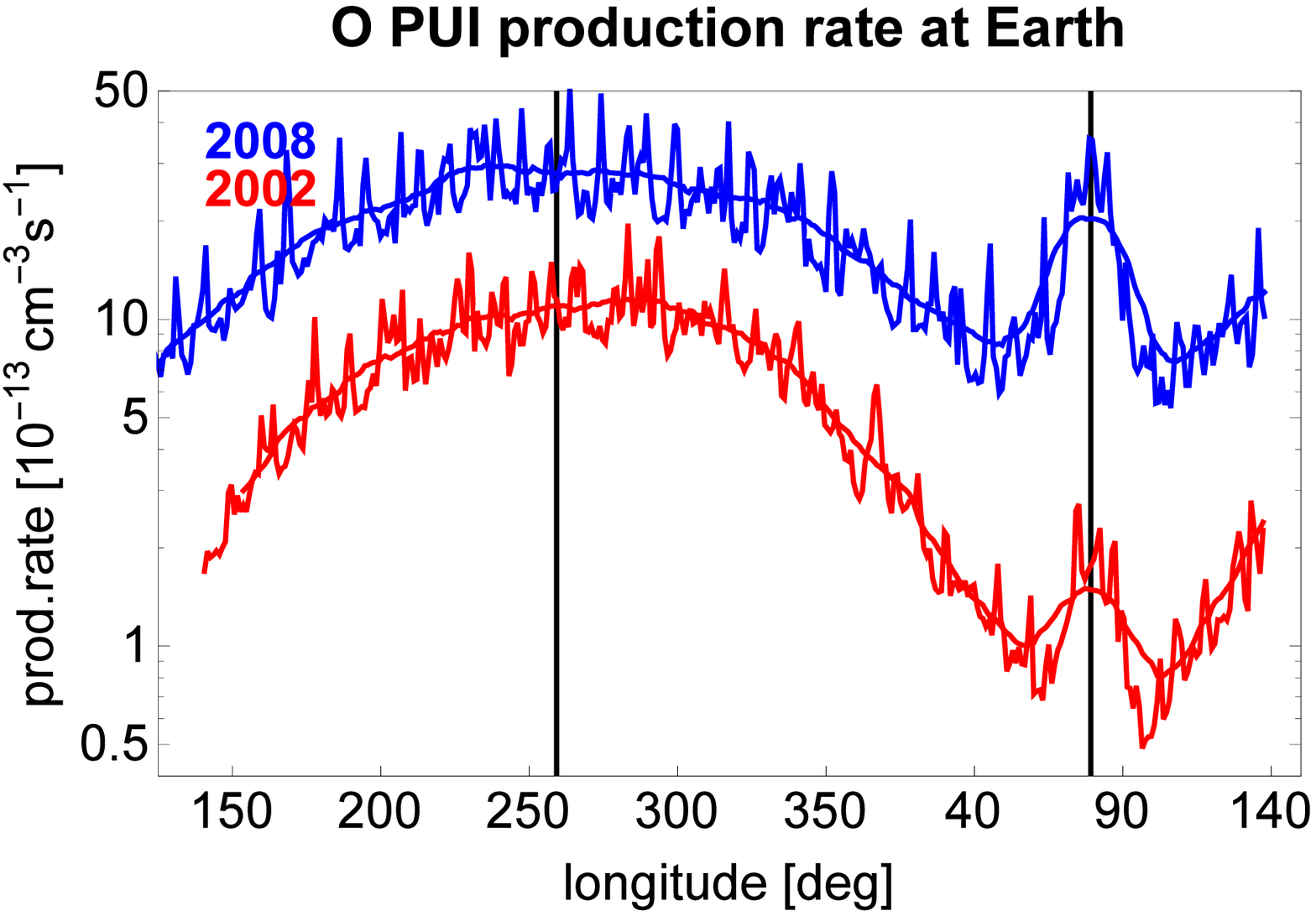} \\
	\includegraphics[scale=0.3]{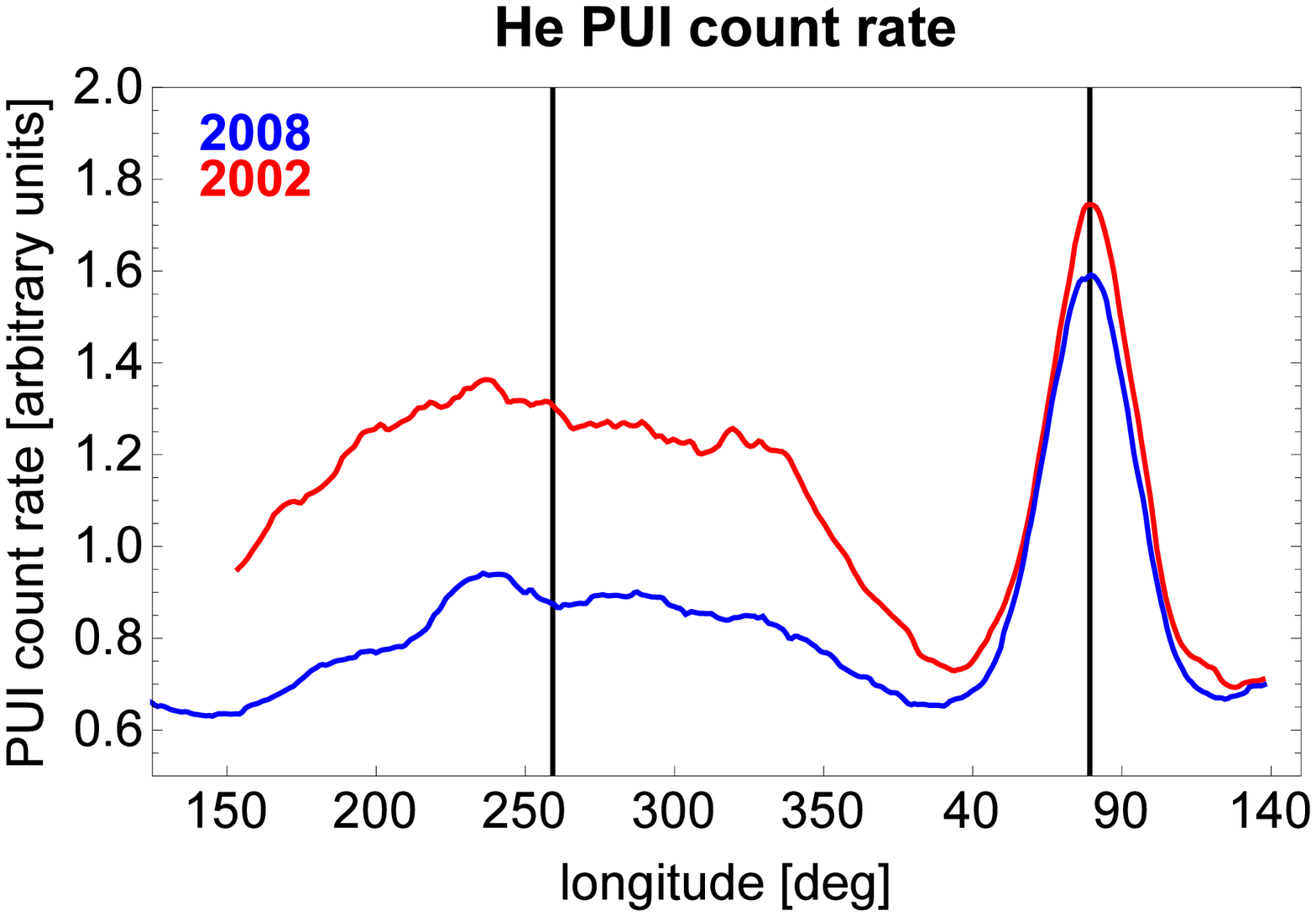} &	\includegraphics[scale=0.3]{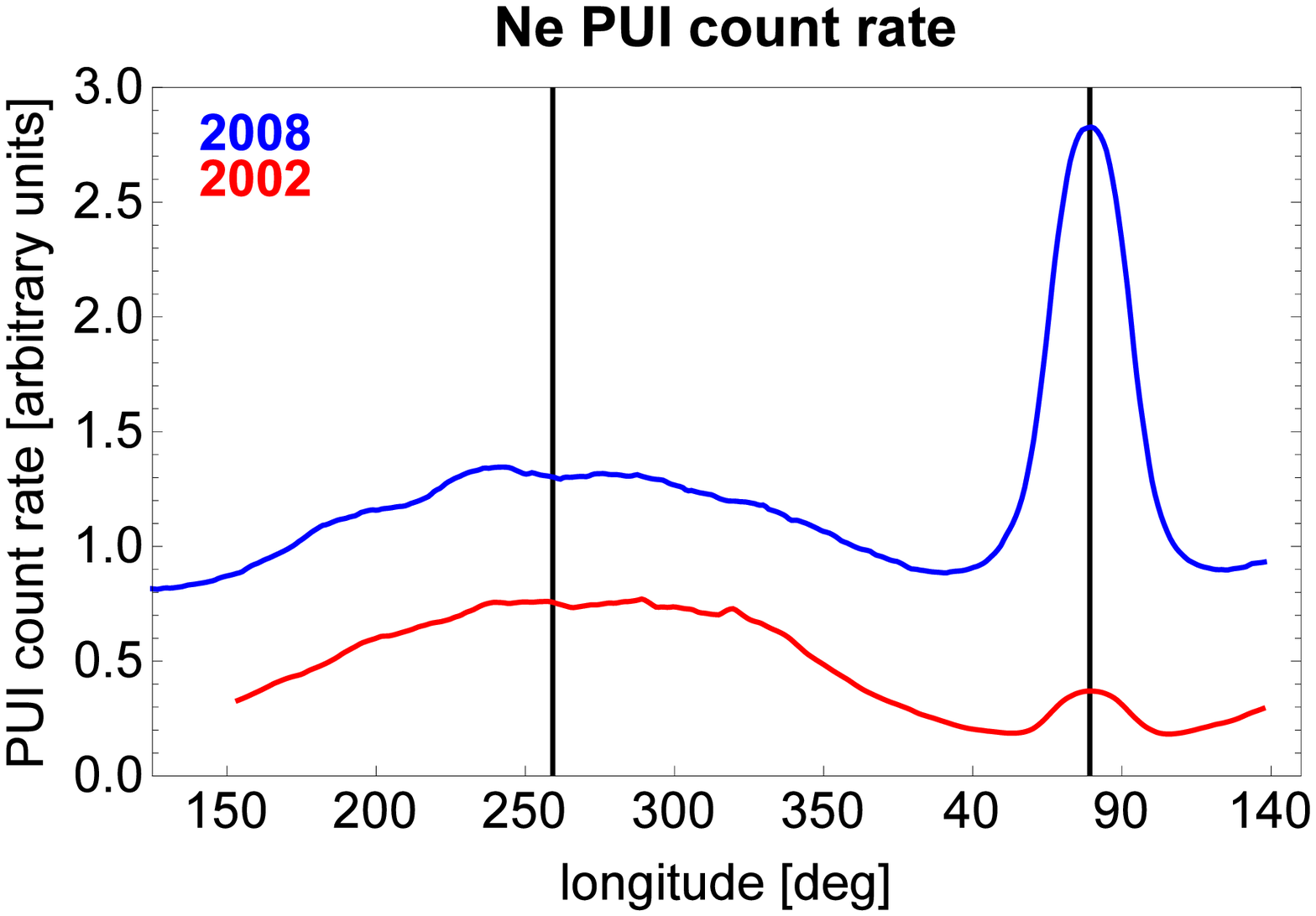} & \includegraphics[scale=0.3]{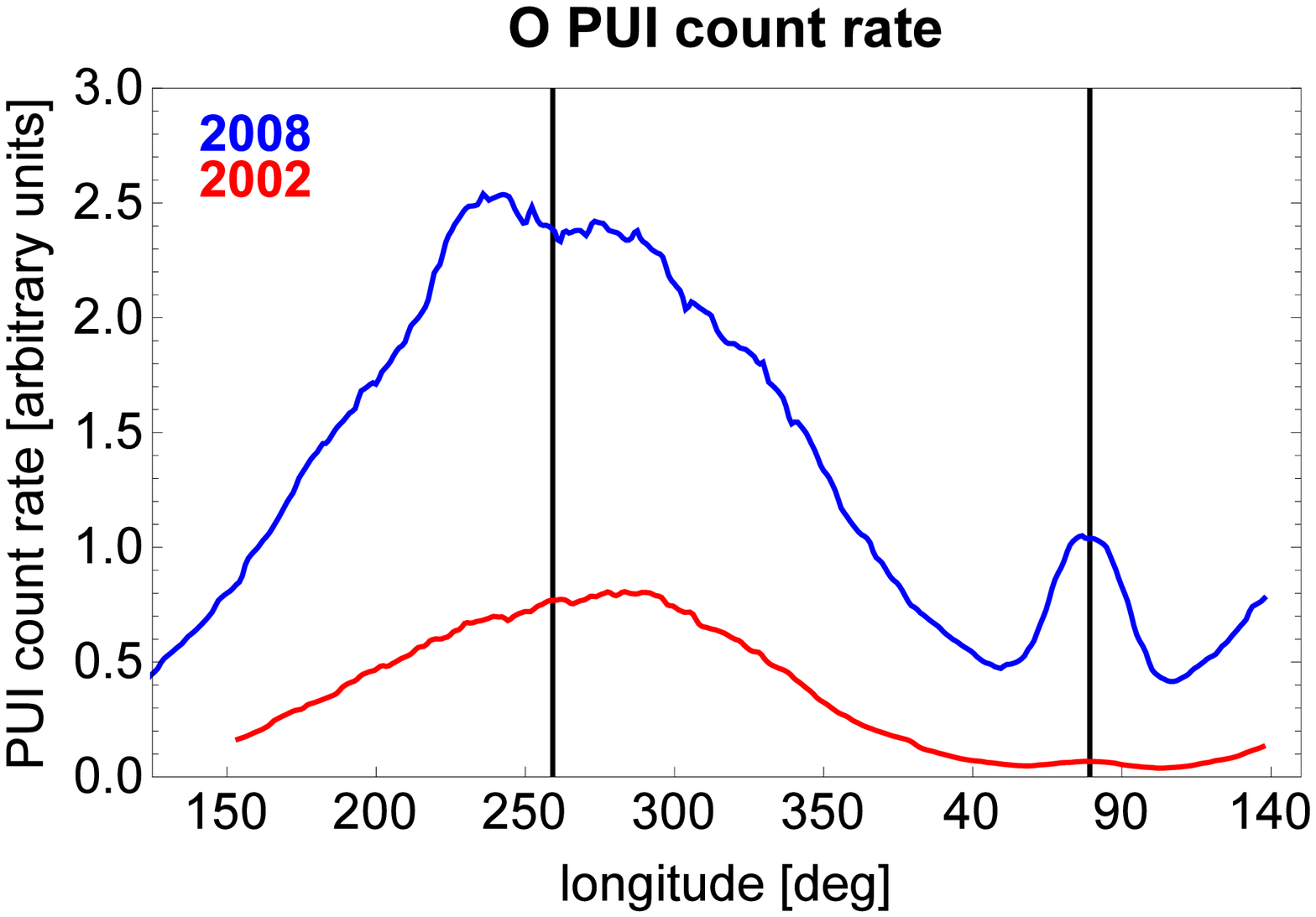} \\
	\end{tabular}
	\caption{Top row: the ISN He, Ne, and O densities. Middle row: the instantaneous local PUI production rates at the Earth's locations for He, Ne, and O, calculated based on Equation~\ref{eq:ProdRate} (daily values -- highly variable and a 27 d moving average). Bottom row: simulated count rates calculated over the range in $\wSC$ based on Equation~\ref{eq:puiCntRate} and normalized by the average value over the whole time interval studied for each species separately, given in arbitrary units. All lines are shown for 2002 (red, solar maximum) and 2008 (blue, solar minimum). The vertical bars indicate the ecliptic longitudes of the upwind and downwind directions of the flow of the ISN gas in the source region assumed in the calculation of the ISN densities.}
	\label{figProdRateTotalPartialFlux}
	\end{figure*}

The time series of the instantaneous local PUI production rates at Earth for He, Ne, and O in 2002 and 2008, calculated according to Equation~\ref{eq:ProdRate}, are shown in the centre row of Fig.~\ref{figProdRateTotalPartialFlux} and the corresponding ISN densities are shown in the upper row of this figure. The daily production rates vary because of fluctuations in the ionization rates (see also the faint lines in the left-hand panel of Fig.~\ref{figIonRatesBetaTot}). In the further analysis, we smoothed the resulting time series by a 27 d moving average to not bias the results by the rapidly varying daily changes and to track the general modulation. The production rates reflect the general shape of the ISN density along the Earth's orbit, in particular the presence or absence of the crescent and cone.  

The flux of PUIs at the distance $R$ from the Sun for a given moment in time can be calculated as an integral of the local production rate along the radial path between the Sun at $r_0$ and $R$ \citep{rucinski_etal:03}:
	\begin{equation}
	F\left(R \right)=\frac{1}{R^2}\int_{r_0}^{R}{S(r){r}^2\mathrm{d}r}=\frac{1}{R^2}\int_{r_0}^{R}{n(r)\beta(r){r}^2\mathrm{d}r}.
	\label{eq:puiFluxGeneral}
	\end{equation}
Note that this form differs from the one frequently used in PUI studies in the treatment of the local ionization rates $\beta$, where typically $\beta(r)=\beta_1\left(\frac{r_1}{r}\right)^2$, with the ionization rate $\beta_1$ at 1~au and $r_1=1$~au. We did not make this simplification (see Section~\ref{sec:Models}) and included the entire radial dependence of $\beta$ self-consistently. The total PUI flux cannot be observed in practice because of the limited energy range of detectors and competition with solar wind ions. Instead, the energy flux density or a portion of the distribution function which typically includes the PUI cut-off is collected. Generally, PUI detectors measure the distribution function of the solar wind and PUI populations as a function of energy per charge in fixed steps and in a restricted FoV. The registration of the energy per charge allows for an independent distinction between the solar wind and PUIs. Because the PUI speed is tied to the solar wind speed (i.e., on average, there is no streaming of the PUI population relative to the solar wind), it is convenient to express the PUI energy as a fraction of the bulk solar wind speed $w^2=\left( \vPUI/\vSW \right)^2$. The integration of the PUI flux is carried out over a fixed interval in the $w$-space that includes the PUI cut-off at $w\sim 2$ in the observer frame, and over the sensor FoV. This approach enables to separate the interstellar PUIs from the inner source PUIs, which are expected close to the Sun, and from the solar wind ions \citep[e.g.][]{gloeckler_geiss:01a, gloeckler_etal:04b, drews_etal:12a, berger_etal:15a, drews_etal:15a, taut_etal:15a}. 

In the present analysis, we modelled the PUI count rate by integrating the production rate of PUIs in the $w$ domain, neglecting the instrumental factors (i.e., real FoV, energy response function). We calculated the PUI count rate in the Sun-centred frame in ecliptic coordinates as seen by a stationary observer at the Earth's distance from the Sun. We followed the classical approach on PUI modelling, assuming instantaneous pitch angle scattering and thus a spherically symmetric PUI distribution in the solar wind frame. We considered adiabatic cooling of the PUI population between their creation and detection and neglected transport- and magnetic-field effects \citep[e.g.][]{fichtner_etal:96a, chalov_fahr:98b, chalov_fahr:99a, fahr_fichtner:11a}. For each time $t$ and distance from the Sun $r$, we calculated the normalized PUI speed in the spacecraft frame $w_{\mathrm{sc}}$ using the relation:
	\begin{equation}	
	\wSC\left(r,t\right)=\frac{\vPUIsc\left(r,t\right)}{\vSW\left(R,t\right)}
	\label{eq:wDef}
	\end{equation}
with $\vPUIsc\left(r,t\right)$ being the speed of the PUI created at distance $r$ from the Sun in the spacecraft frame and $\vSW\left(R,t\right)$ the solar wind speed measured at the time and distance of detection. We assumed that the PUIs flow with the solar wind speed that was registered by the detector at the moment of detection. A more sophisticated approach would require magnetohydrodynamic (MHD) modelling to trace the solar wind flow that carries the PUIs, which is beyond the scope of this paper. We made this simplification because of lack of solar wind observations inside the Earth's orbit and we used the daily time series of in-ecliptic solar wind at the Earth's distance to the Sun, smoothed by a 27 d moving average. This is a reasonable assumption because the solar wind speed is constant for the distance range interesting for the PUI observations.

In the solar wind frame, the PUI speed at the distance of their creation is the magnitude of the vector sum of the  interstellar flow and solar wind velocity. As discussed by \citet{mobius_etal:15c}, it is sufficient to only consider the radial components, which yields:
	\begin{equation}
  \vPUIsw(r) =  \vSW(R) - \vrad(r),
 	\label{eq:vPUIsw} 
 	\end{equation}
where $\vrad$ is the radial component of the ISN bulk velocity vector, as defined, e.g., by \citet{bzowski_etal:97}. At the observer distance $R$ the speed of PUIs that move exactly radially outward is the sum of the solar wind speed and the PUI speed in the solar wind frame ($\vPUIsw\left(r\right)$) after adiabatic cooling from their creation at distance $r$, according to equation 3.2 in \citet[][see also \citep{mobius_etal:99}]{mobius_etal:95}:
	\begin{equation}
	\vPUIsc\left(R\right) = \vPUIsw\left(r\right) \left(\frac{R}{r}\right)^{-\gamma}+\vSW\left(R\right),
	\label{eq:vPUIsc}
	\end{equation}
where $\gamma$ is the cooling index (equal to $\frac{2}{3}$ for adiabatic cooling), and $R$ is the observer distance (equal to $\Rd$ in this study).
% radial speed
	\begin{figure}
	\centering
	\includegraphics[width=\columnwidth]{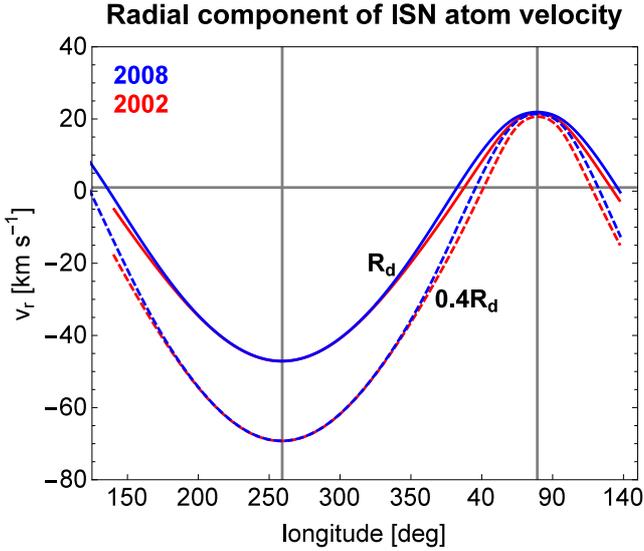}\\
	\caption{The radial component of the ISN He bulk flow velocity vector along the Earth's orbit for 2002 (red lines) and 2008 (blue lines) at $\Rd$ (solid lines) and $0.4\Rd$ (dashed lines). The vertical bars indicate the longitudes of the upwind and downwind directions adopted in the calculations.}
	\label{figSpeedRadial}
\end{figure}
% wSC for He and % w energy cut off for ISN PUIs
	\begin{figure}
		\begin{center}
		\begin{tabular}{c}
		\includegraphics[width=\columnwidth]{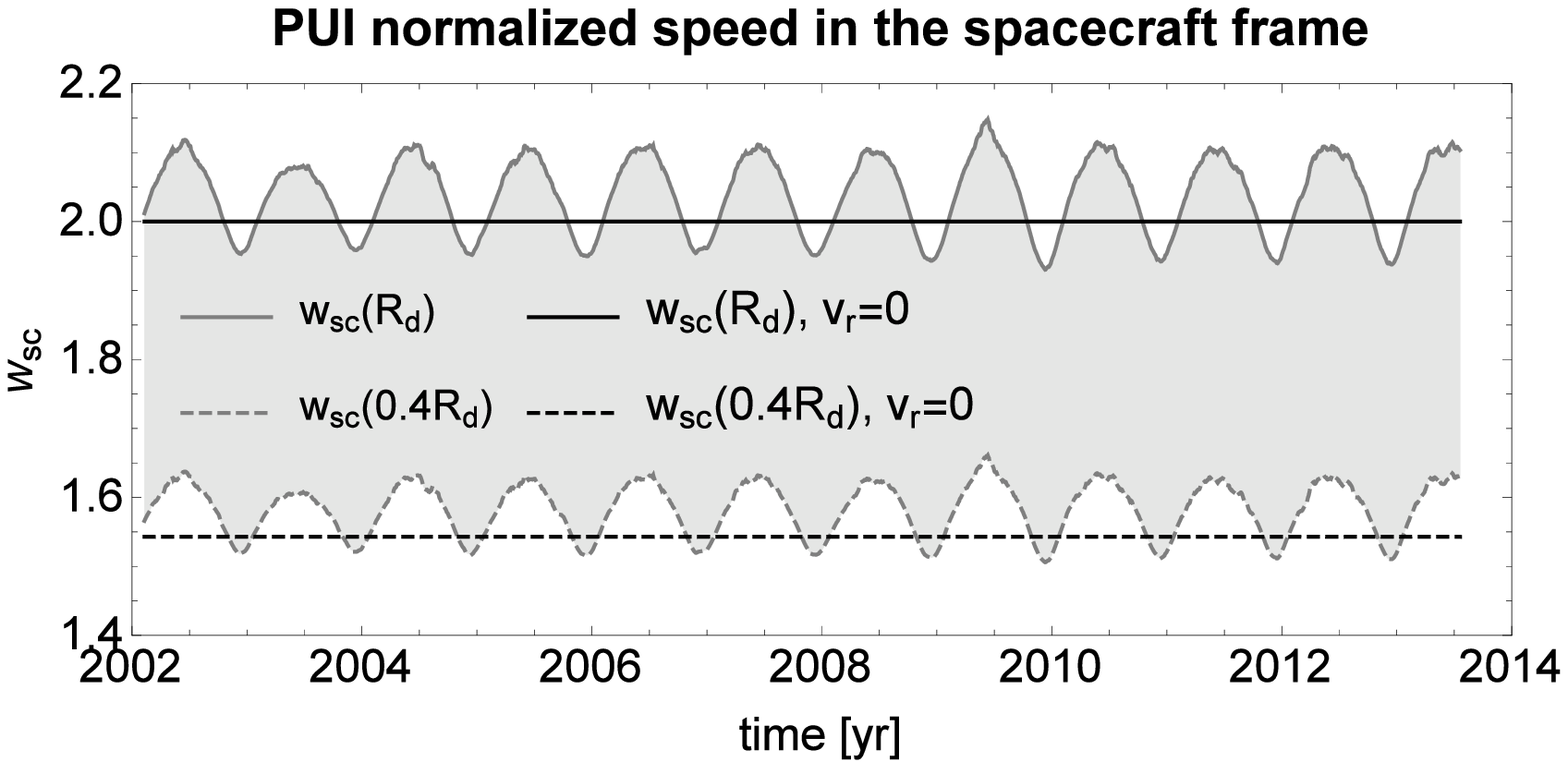}\\
		\includegraphics[width=\columnwidth]{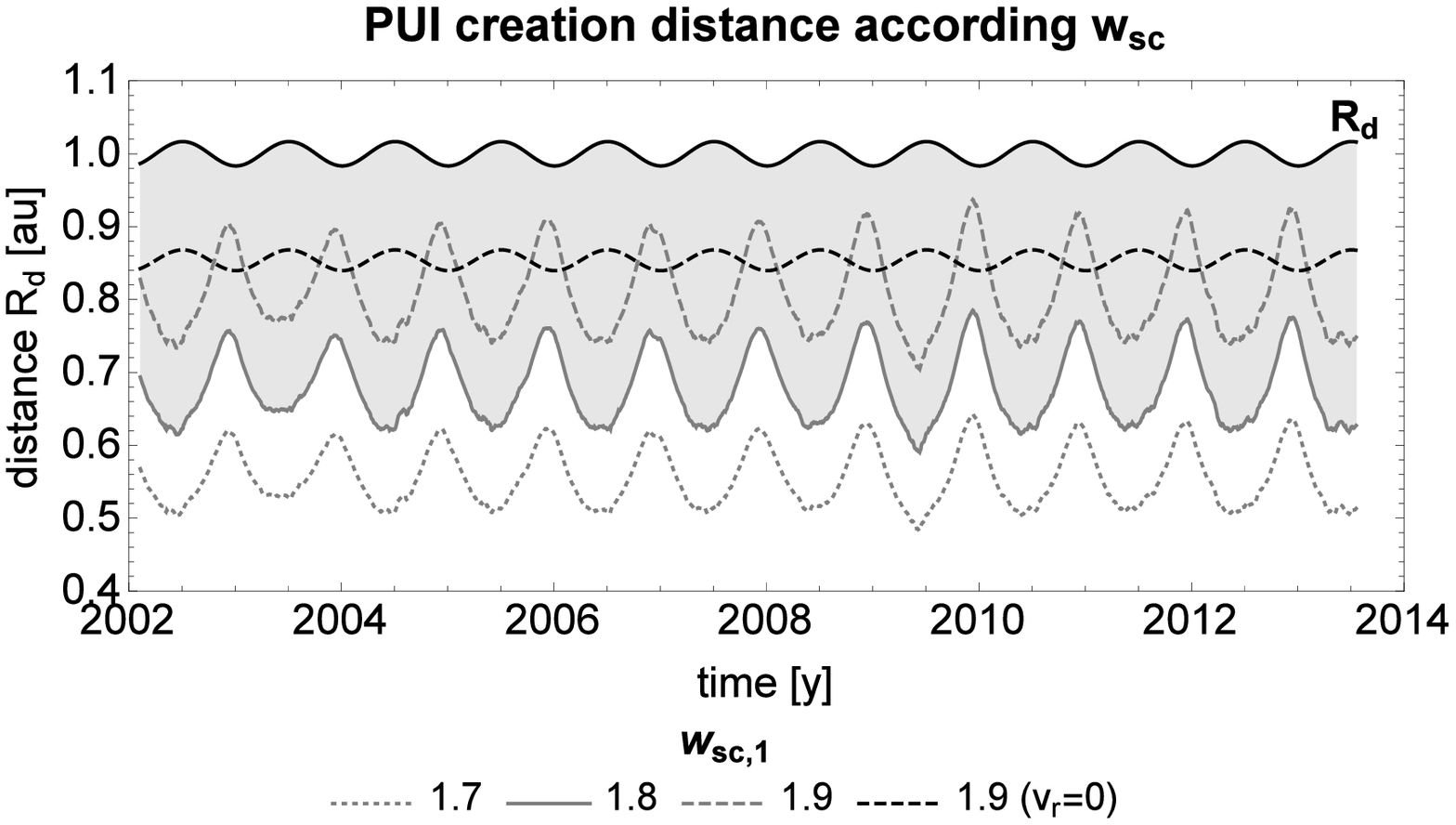}\\
		\end{tabular}
		\end{center}
		\caption{Top panel: time series of the normalized PUI speed $\wSC$ calculated from the formula in Equation~\ref{eq:wSC}, for PUIs detected at $\Rd$, but created at $0.4\Rd$ and $\Rd$ taking adiabatic cooling into account. Presented are series for He, but they are very similar for Ne and O. The horizontal lines illustrate the case for which the radial velocity of ISN atoms at the moment of ionization is neglected. Bottom panel: the distance of the origin of PUIs calculated for different values of the lower boundary $w_{\mathrm{sc,1}}$ in the integral in Equation~\ref{eq:puiFlux2} based on Equation~\ref{eq:r-wSC}, shown for the longitude of the Earth changing with time. The black solid line $\Rd$ represents the longitude-dependent Earth's distance to the Sun.} 
	\label{figWsc2}
	\end{figure}

Fig.~\ref{figSpeedRadial} illustrates the radial component ($\vrad$) of the mean velocity vector of the ISN~He flow distribution at $\Rd$ and $0.4\Rd$ for solar minimum and solar maximum. The variation of the radial speed with ecliptic longitude along the Earth's orbit is almost identical for He, Ne, and O, with only minor differences around the cone. The extrema around the upwind and downwind directions are well pronounced. The radial speed depends weakly on the phase of solar cycle, with differences only close to the cone because of the differential action of ionization \citep{bzowski_etal:97}. Ionization more readily eliminates the slower atoms from the distribution, which results in an effect of acceleration (change in velocity vector) of the whole ISN population. In particular, the mean velocity is a superposition of the so-called direct and indirect beams, with the indirect beam more attenuated during solar maximum than during solar minimum. This mechanism is the most efficient for ISN He, because its thermal speed is the largest. Interestingly, the radial ISN speed at the cone is almost independent of the solar distance. The distance-related differences are much more pronounced for the crescent (they reach $\sim20$~km~s$^{-1}$). 

For observations carried out at the distance $\Rd$ from the Sun, the normalized PUI speed in the spacecraft frame $\wSC=\vPUIsc/\vSW$ is modulated by the injection speed and modified by adiabatic cooling. This is illustrated in the top panel of Fig.~\ref{figWsc2}, where are shown the normalized speeds of PUIs detected at $\Rd$, but created at $\Rd$ and $0.4\Rd$. The quasi-periodic variations are due to the non-zero injection speed as shown in Fig.~\ref{figSpeedRadial}, and the small-scale fluctuations due to the daily variations in solar wind speed. The macroscopic difference between $\wSC(\Rd)$ and $\wSC(0.4\Rd)$ is due to the adiabatic cooling of the PUIs over the distance $0.4\Rd$--$\Rd$ on one hand and the lack of this cooling for the PUIs created at $\Rd$. When the non-zero PUI injection speed is neglected, the normalized speed of PUIs created at $0.4\Rd$ and $\Rd$ would differ only due to adiabatic cooling, without the yearly modulation, as illustrated by the horizontal lines in the upper panel of Fig.~\ref{figWsc2}.

For the PUIs created at $\Rd$ from the Sun, the expected value of $\wSC$ exceeds 2 in the upwind hemisphere and decreases below 2 close to the cone. As evident from Equations~\ref{eq:puiFluxGeneral} and \ref{eq:puiFlux2}, integration of the PUI flux can be carried out either in the domain of solar distances $r$ or, equivalently, in the domain of normalized PUI speeds $\wSC$, where $\wSC$ is tied to $r$ by Equation~\ref{eq:r-wSC}. Under assumption of zero injection velocities selecting a fixed integration range in the $\wSC$ domain is equivalent to selecting a fixed interval in the $r$ domain which does not vary with ecliptic longitude. However, when the assumption of the injection speed is removed, keeping fixed integration range in $r$ results in the necessity to appropriately adjust the integration boundaries in the $\wSC$ domain. This is illustrated in the top panel of Fig.~\ref{figWsc2}, where for a fixed interval from $0.4\Rd$ to $1\Rd$ we show the corresponding boundaries in the $\wSC$ domain. Frequently in experimental practice, the PUI count rate is obtained by integration over a fixed range in the $\wSC$ domain. This approach will be exercised in the remainder of this paper. Adopting a fixed range of integration in the $\wSC$ domain results in admitting the PUIs created over range of heliocentric distances that varies with ecliptic longitude of the observer, as illustrated in the bottom panel of Fig.~\ref{figWsc2}.   

The calculation of the PUI flux in the domain of normalized PUI speed in the spacecraft frame requires a change of variables in Equation~\ref{eq:puiFluxGeneral} from distance $r$ to $\wSC$. The derivation is shown in Appendix~\ref{sec:appendix}. The resulting formula for the PUI flux is the following:
	\begin{equation}
	F=\alpha R \int\limits_{w_{\mathrm{sc,1}}}^{w_{\mathrm{sc,2}}}{ S\, \wSW^{-3\alpha}\left(\wSC-1\right)^{3\alpha-1}\mathrm{d}\wSC} ,
	\label{eq:puiFlux2}
	\end{equation}
with $\alpha=\frac{1}{\gamma}$, and $R=\Rd$. 

PUIs are typically measured using a sensor that includes an electrostatic analyzer. The quantity registered by the sensor is the count rate, which is linked to the phase space density of PUIs $f(v)$ (assuming an isotropic distribution in the solar wind frame) and the speed $v$ of the sampled particles in the observer frame. The formula for PUI count rates in the $\wSC$ domain is the following (for details see Appendix~\ref{sec:appendix}):
	\begin{equation}
	C_p = \frac{1}{2}\alpha R A \int\limits_{w_{\mathrm{sc,1}}}^{w_{\mathrm{sc,2}}}S\,\wSW^{-3\alpha} \left(\wSC-1\right)^{3\alpha-1}\wSC \mathrm{d}\wSC,
	\label{eq:puiCntRate}
	\end{equation}
where $A$ is a proportionality constant depending on details of the instrument.

We calculated the expected PUI count rate using Equation~\ref{eq:puiCntRate} assuming a constant lower integration boundary $w_{\mathrm{sc,1}}$ given by one of the values $w_{\mathrm{sc,1}}=1.7$, 1.8, or 1.9. For the upper integration boundary $w_{\mathrm{sc,2}}$ we evaluated Equations~\ref{eq:wSC} at $r=R=\Rd$, which leads to $w_{\mathrm{sc,2}} = 2-\frac{\vrad}{\vSW}$ and includes the variation with ecliptic longitude. The variable upper integration limit is equivalent to choosing the high fixed value used by \citet{drews_etal:12a} because the phase space above the PUI cut-off is empty. 

\begin{figure*}
	\centering
	\includegraphics[scale=0.8]{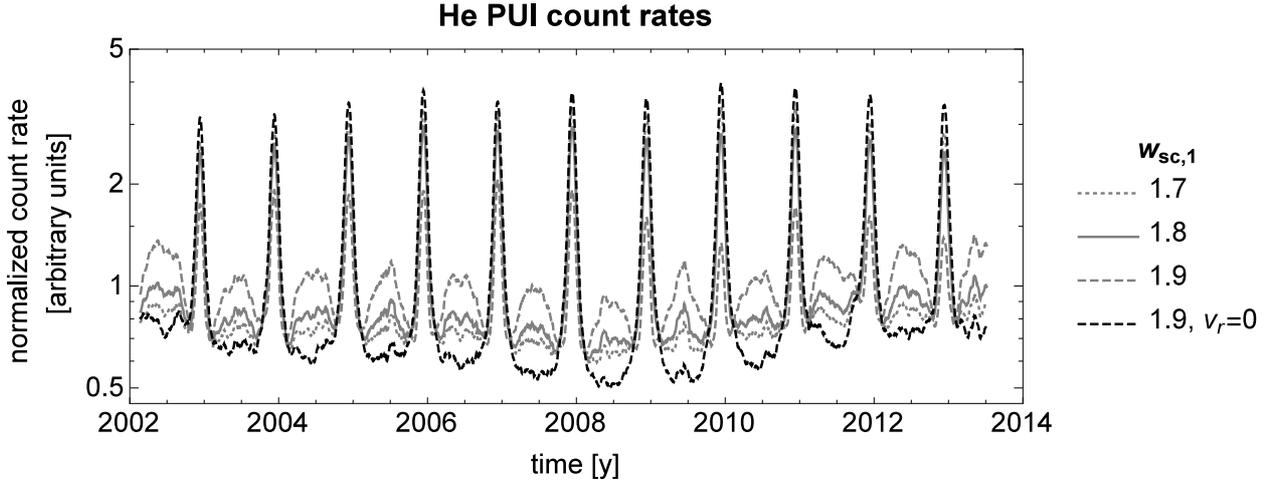}\\
	\caption{Time series of simulated PUI count rates for He at Earth, calculated based on Equation~\ref{eq:puiFlux2}, with $w_{\mathrm{sc,1}}=\left(1.7,\,1.8,\,1.9\right)$ and normalized by the averaged value for the studied time series, given in arbitrary units. The black dashed line represents the case with the radial speed $\vrad$ ignored in the integration of the PUI count rates.}
	\label{figCntRateHeW}
	\end{figure*}

Fig.~\ref{figCntRateHeW} illustrates the resulting modulation of the expected He~PUI count rate calculated for various values of $w_{\mathrm{sc,1}}$ in Equation~\ref{eq:puiCntRate}. The differences are most pronounced in the upwind part of the orbit. The crescent forms in all solar activity conditions for $w_{\mathrm{sc,1}}=1.9$, but is less pronounced during solar minimum for $w_{\mathrm{sc,1}}=1.7$ and $w_{\mathrm{sc,1}}=1.8$. The contrast between the cone and the crescent peaks varies with the change $w_{\mathrm{sc,1}}$. It is higher for the lower limit and decreases with an increase in $w_{\mathrm{sc,1}}$. The count rate series are flat at the upwind side of the Earth's orbit when calculated with zero injection speed. In this case, the downwind cone is the dominant feature in the annual pattern, and its magnitude depends weakly on the phase of solar activity. We conclude that the factor responsible for the creation of the He PUI crescent is the non-zero PUI injection speed.
% partial flux for He and O

For the Ne and O PUI count rate, however, the crescent exists regardless of the choice of $w_{\mathrm{sc,1}}$, but the cone as seen from a detector in the Earth's orbit almost disappears for higher values of $w_{\mathrm{sc,1}}$ in the case of ISN~O. The visibility of the cone in the Earth's orbit strongly depends on solar activity. The more active the Sun, the higher the ionization rate, and the fainter the cone observed from the ecliptic plane. Thus, the most favorable times to analyse the PUI cone are during solar minimum. 

Based on these findings, we chose $w_{\mathrm{sc,1}}=1.8$ as the lower integration boundary for Ne and O, and $w_{\mathrm{sc,1}}=1.9$ for He for further analysis. The resulting time series of the PUI count rate is shown in the lower row of Fig.~\ref{figProdRateTotalPartialFlux} for 2002 and 2008. In contrast to the ISN density (top panels) and instantaneous local production rate in the Earth's orbit (middle panels), the crescent for the He~PUI count rate is now visible for both years. Also for the other species, both structures, cone and crescent, are present, with the cone strongly reduced for the O PUI count rate during solar maximum. 

\section{Derivation of the ISN flow longitude from the PUI count rate}
\label{sec:fits}
The longitude of the ISN inflow is often determined from the positions of the peak of the PUI cone and crescent (e.g., \citealt{gloeckler_etal:04b, drews_etal:12a}). We looked for the positions of the peaks by fitting a Gaussian function to the crescents and cones for individual years in the time series in Fig.~\ref{figProdRateTotalPartialFlux}: the density of ISN atoms, the instantaneous local production rate of PUIs, and the PUI count rate. The Gaussian function was fitted to the 27 d moving average series of these quantities (except for the ISN density, for which the time series are smooth enough). Note that smoothing by a 27 d moving average broadens the width of the cone more for the heavier species, for which the cone is narrow (compare strongly varying and smooth lines in middle panel of Fig.~\ref{figProdRateTotalPartialFlux} for Ne and O), but this effect does not bias the determination of the location of the peak. The fits were performed to data selected for the following ranges in longitude: $\left(65\degr-95\degr, 70\degr-90\degr, 71\degr-87\degr \right)$ for the ISN He, Ne, O density cone, respectively; $\left(160\degr-10\degr, 130\degr-30\degr \right)$ for the ISN Ne and O density crescent, respectively. For the local PUI production rate and count rate at Earth, the longitudinal ranges were adopted as follows: $\left( 55\degr-105\degr, 70\degr-90\degr, 65\degr-90\degr \right)$, for He, Ne, O cone, respectively; and $\left( 150\degr-20\degr, 130\degr-20\degr, 110\degr-50\degr \right)$ continuously counted through $360\degr$ for the He, Ne, O crescents, respectively.

\begin{figure*}
	\centering	
	\begin{tabular}{cc}
	\includegraphics[scale=0.4]{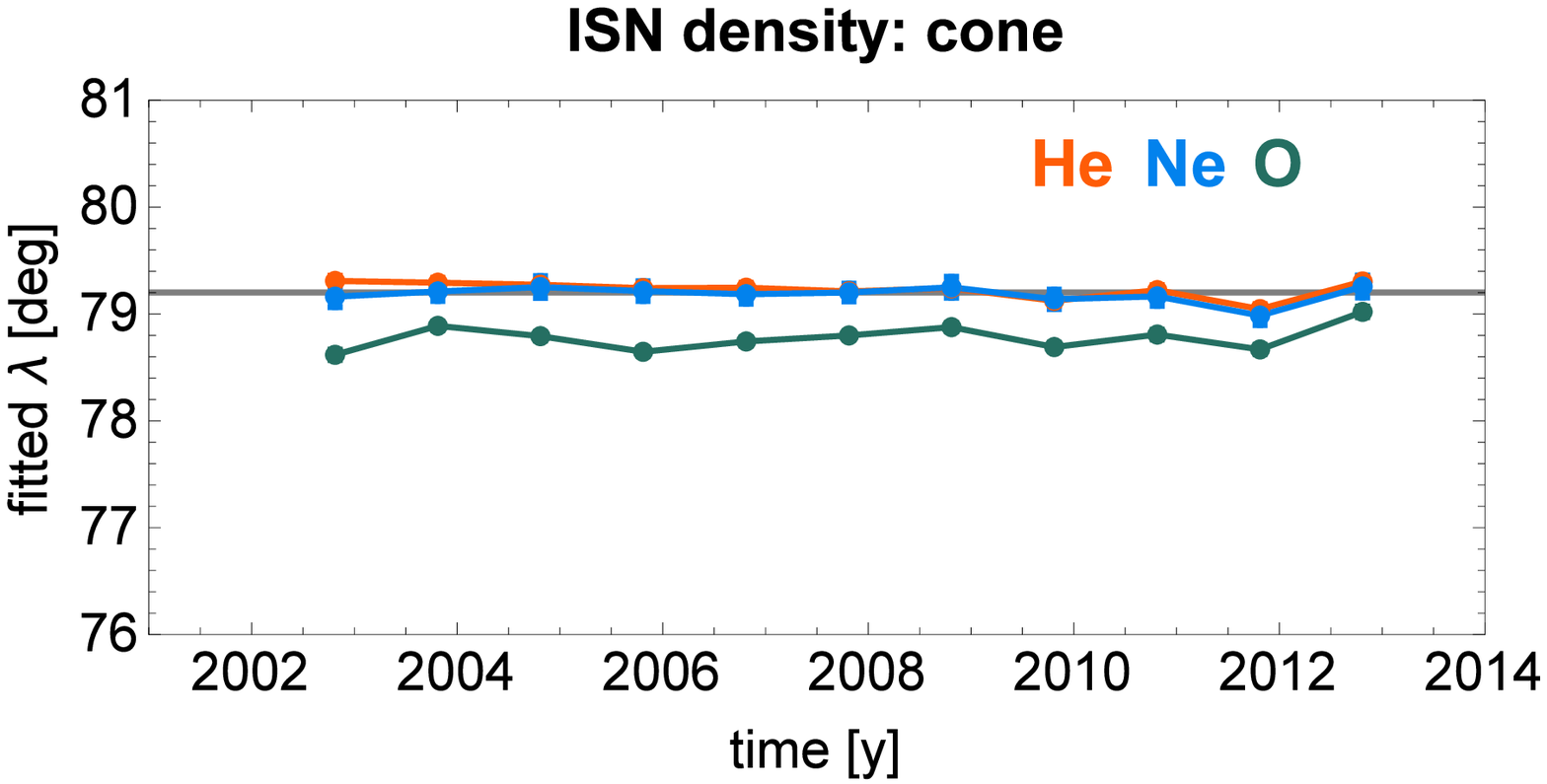} & \includegraphics[scale=0.4]{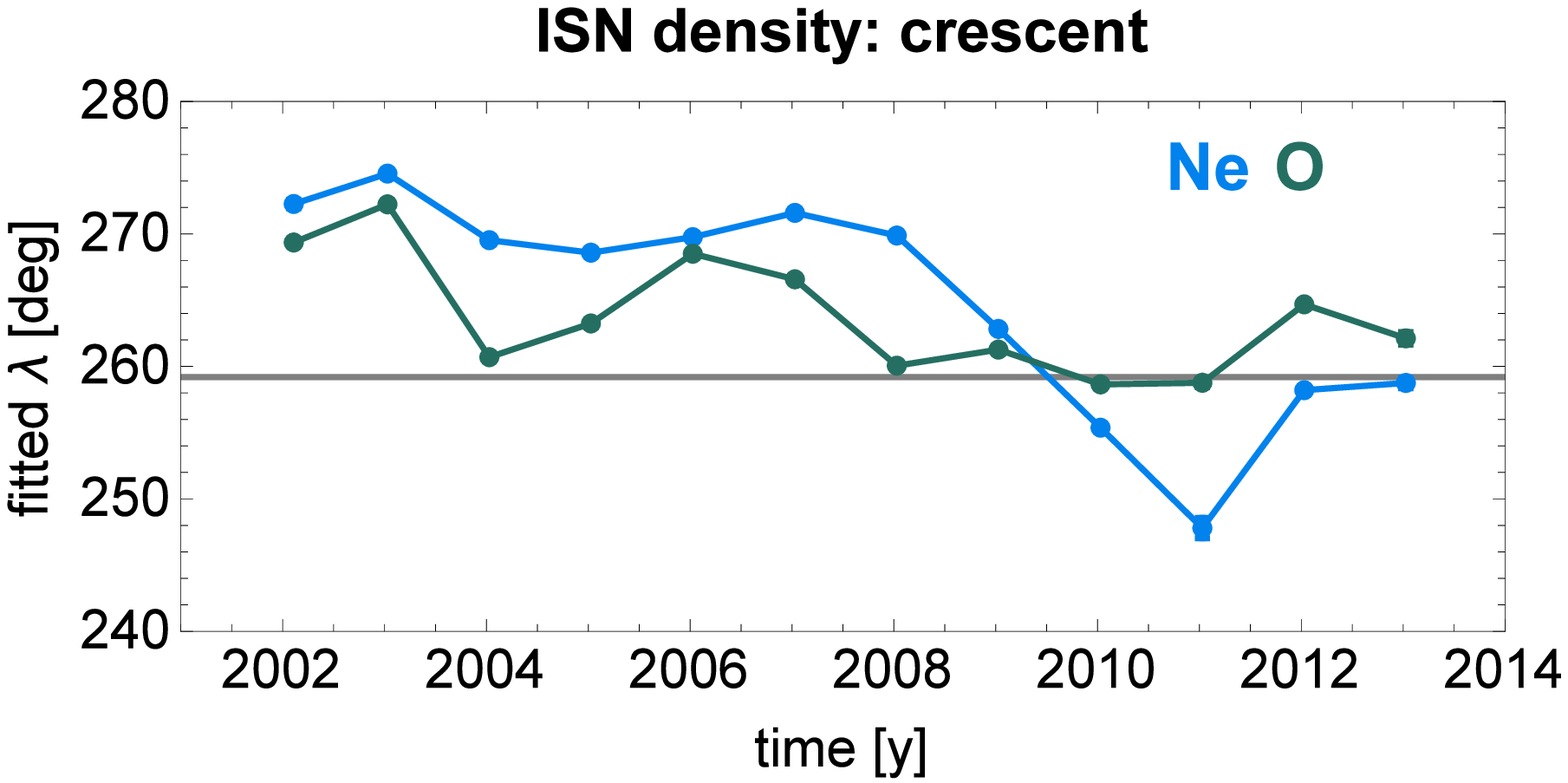} \\
  \includegraphics[scale=0.4]{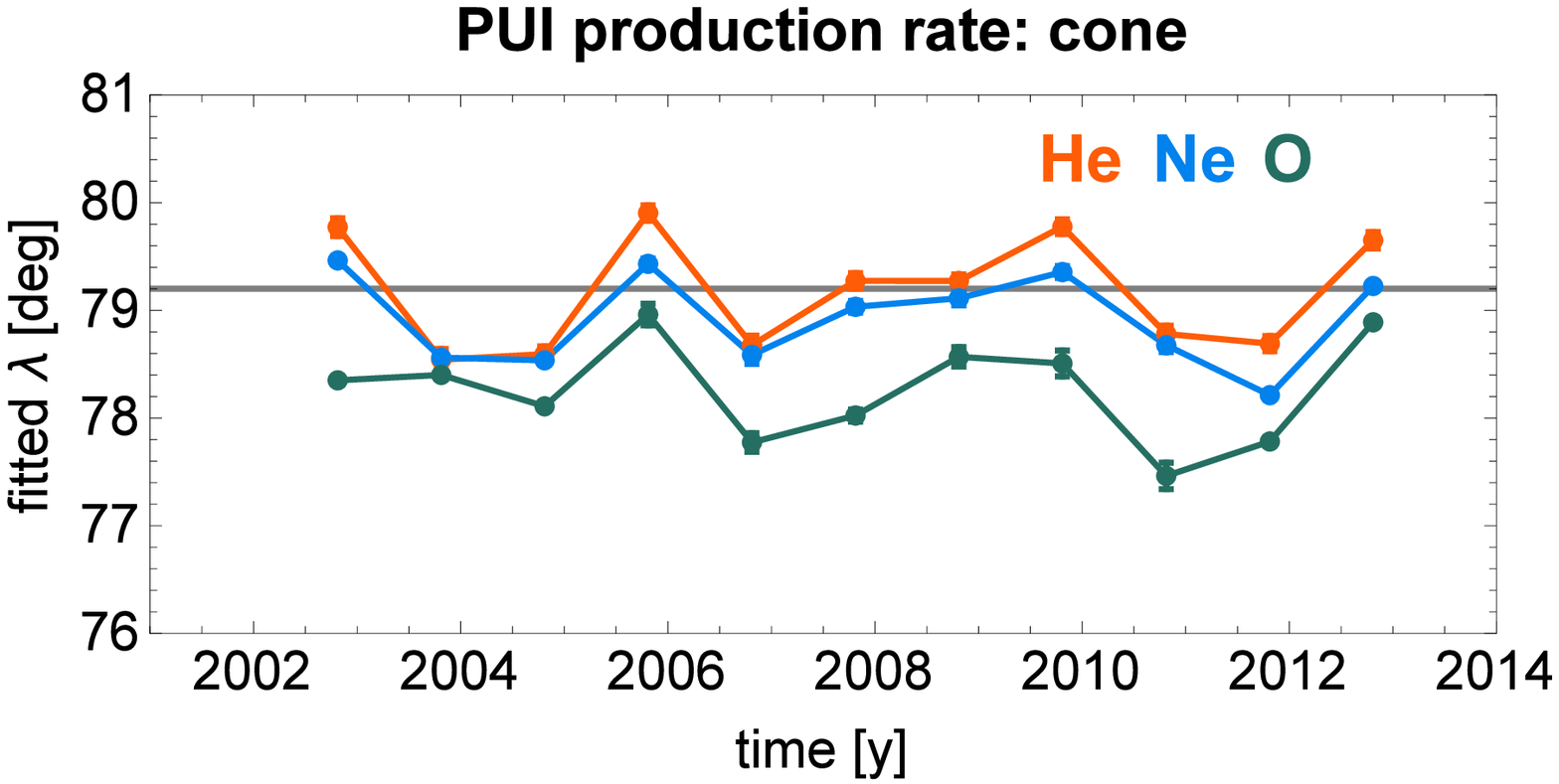} & \includegraphics[scale=0.4]{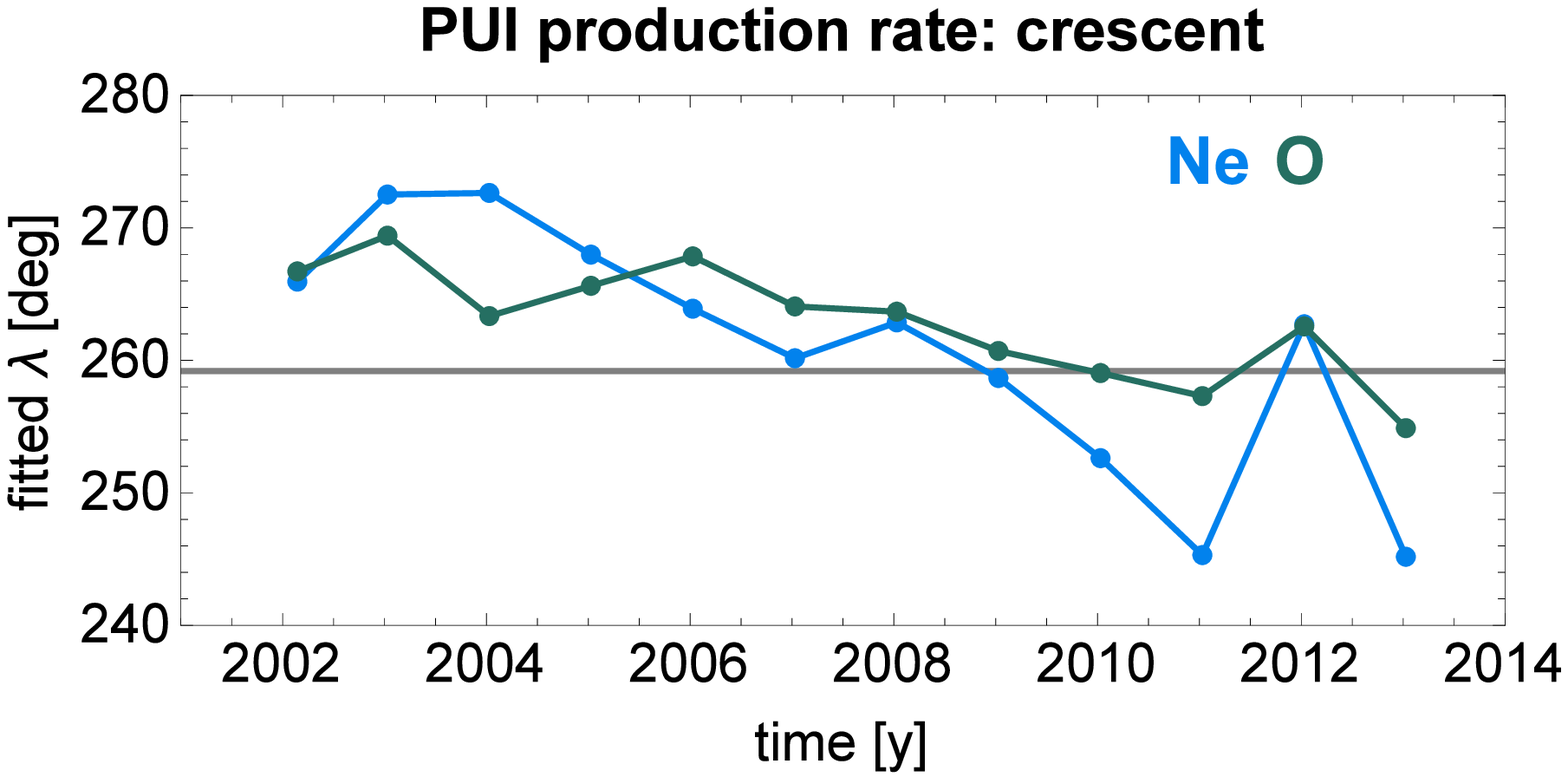} \\
	\includegraphics[scale=0.4]{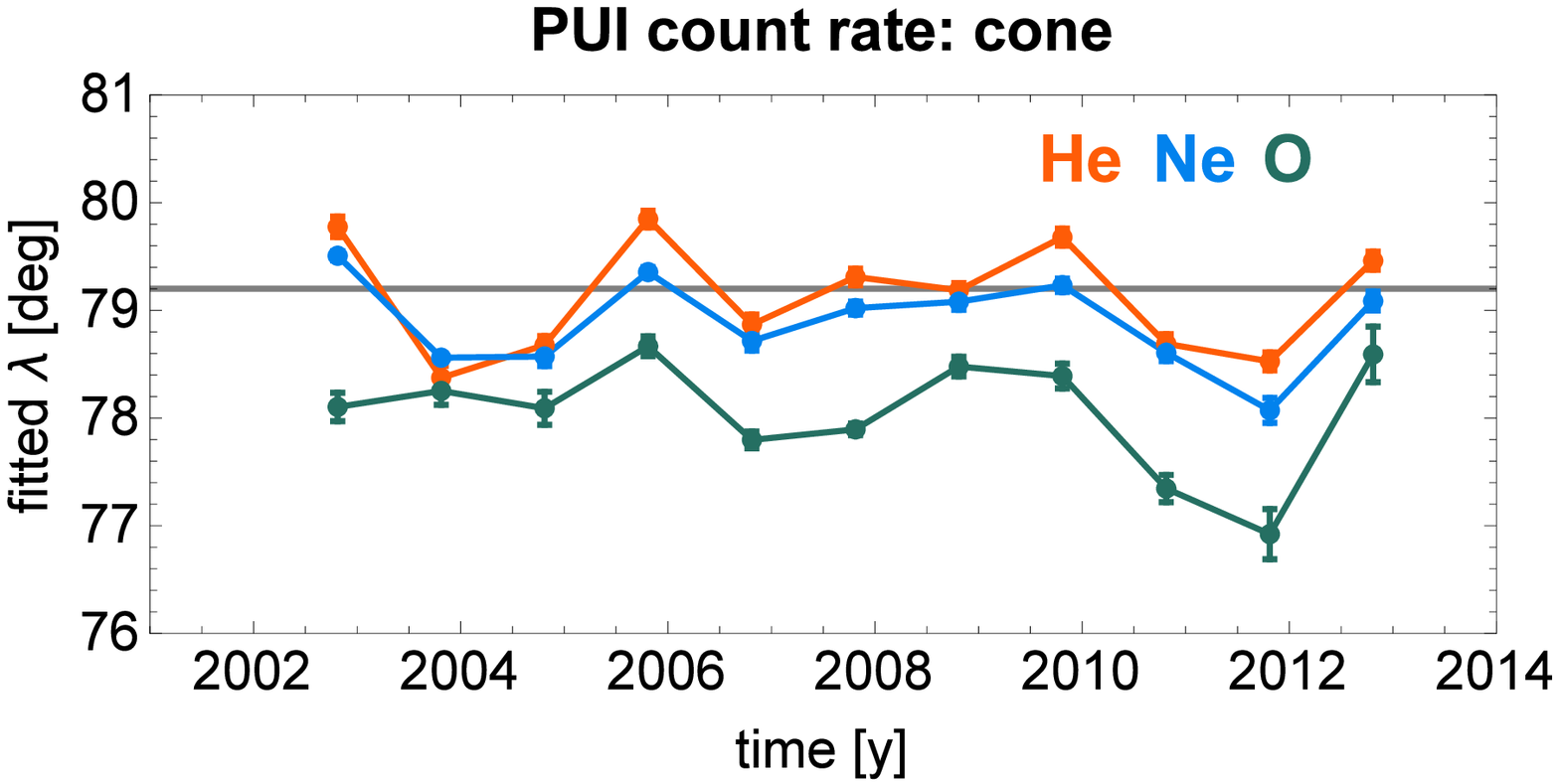} & \includegraphics[scale=0.4]{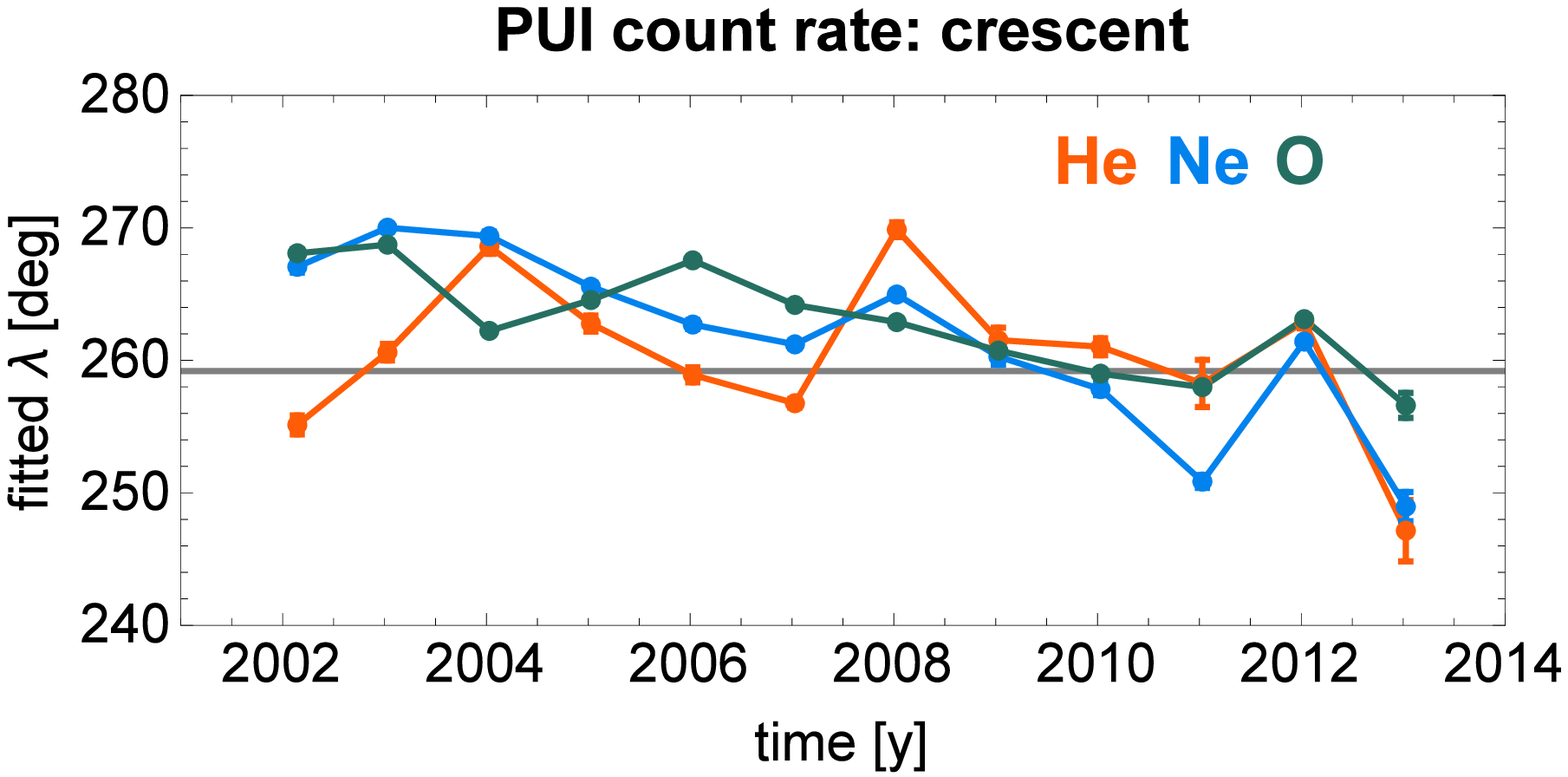} \\
		\end{tabular}
	\caption{Longitude $(\lambda)$ of the peaks of the focusing cone (left-hand panels) and crescent (right-hand panels) from the Gaussian fits to the ISN density (top row) and 27 d moving averages of the local production rate at Earth of PUIs (second row from top) and simulated count rate integrated over normalized PUI speed $\wSC$ (bottom row) for He (orange), Ne (blue), and O (green) for the case of detection at the Earth's orbit ($\Rd$). The horizontal solid line mark the direction of the ISN gas in the source region for upwind ($259.2\degr$) and downwind ($79.2\degr$) sides, assumed in the calculation of densities. The uncertainties of the fits are illustrated by the error bars, in most cases they are smaller than the spatial dimension of the dots.}
	\label{figPeakGaussFit}
	\end{figure*}

\begin{figure*}
	\begin{center}
	\begin{tabular}{cc}
	\includegraphics[scale=0.4]{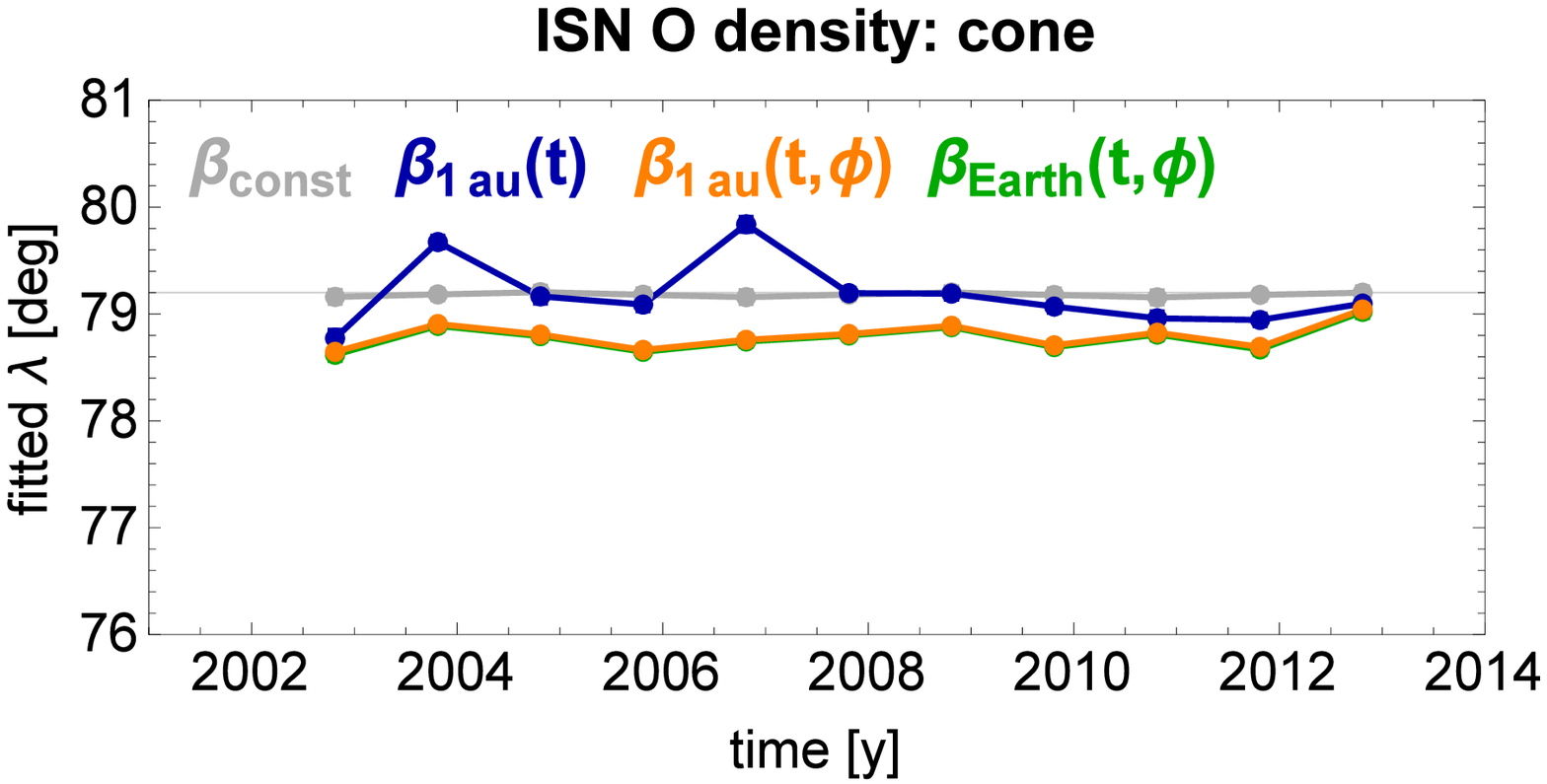} & \includegraphics[scale=0.4]{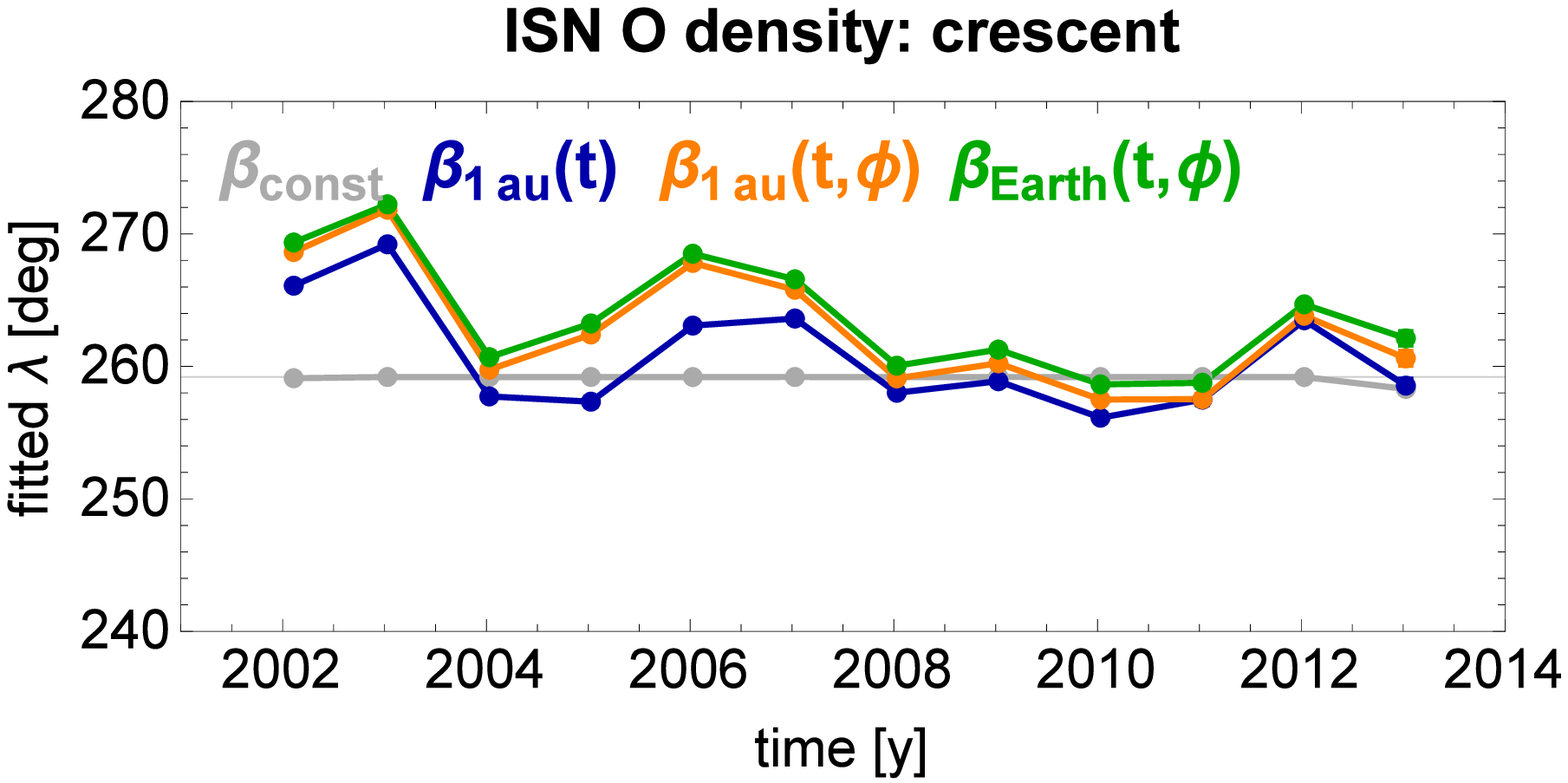} \\
	\end{tabular}
	\end{center}
	\caption{The analysis of factors potentially responsible for shifting the peak longitude ($\lambda$) (cone left-hand panel, crescent right-hand panel) for ISN~O density for the cases shown in Fig.~\ref{figDensityOxEarthFull}. The peak positions are determined from Gaussian fits. The colour code is the same as in Fig.~\ref{figDensityOxEarthFull}. Notice that the green line is almost totally hidden under the orange line in the left-hand panel. The uncertainties of the fits are illustrated by the error bars, in most cases they are smaller than the spatial dimension of the dots.}
	\label{figPeakGaussOxExercise}
	\end{figure*}

We analysed expected departures of the positions of the peaks obtained in our model of the PUI count rate observed at Earth from the ISN inflow longitude assumed in the model. Fig.~\ref{figPeakGaussFit} shows the longitudes of the fitted peak positions. The left-hand column panels show the results for the cones and the right-hand column panels show the results for the crescents. The horizontal line in each panel indicates the assumed ecliptic longitude of the ISN gas flow in front of the heliosphere used in the calculation of the ISN density ($79.2\degr$ downwind and $259.2\degr$ upwind). In an ideal case, when modulation of the ISN gas and the PUIs does not exist inside the heliosphere, the fitted longitudes should be exactly these values. In the case when modulation is accounted to the longitude direction derived from the fit to the density of ISN atoms at Earth almost perfectly reproduces the ISN flow longitude for He and Ne and a small systematic shift is found for O. This $\sim1\degr$ shift in the cone position of the ISN~O density is caused by the asymmetry of the solar wind in heliolatitude, reflected in the charge exchange rate (as presented in Fig.~\ref{figPeakGaussOxExercise}). The values obtained from fitting to the cone of the local PUI production rate and PUI count rate show a scatter of $\sim1\degr$ for He and Ne, and a systematic $\sim 1\degr - 2\degr$ downward shift for O. Additionally, the fit is closer to the assumed value for years of solar minimum, when ionization is weaker and varies little. There is a clear year-to-year scatter which does not clearly correspond to the solar activity cycle. Since the density peak varies very little in time, we conclude that the scatter in the peak positions of the quantities other than density is caused by the fluctuations in the PUI production rates and in the solar wind that carry the PUIs. This conclusion is supported by the left-hand panel of Fig.~\ref{figPeakGaussOxExercise}, which illustrates the fitted positions of the O density peak for densities calculated using various assumptions on ionization rates, similarly as it was in the case of Fig.~\ref{figDensityOxEarthFull}. Clearly, the case of a circular orbit and constant ionization produces no shift, switching in the time variations adds the yearly scatter in the peak positions, adding the latitude dependence adds the systematic shift, and the departure from orbit circularity does not modify these conclusions.

The derivation of the longitude of the ISN gas flow from fits to the crescent (right-hand panels of Fig.~\ref{figPeakGaussFit}) is less accurate and can differ from the expected value up to $10\degr$. Based on similar analysis to that just presented for the cone positions, the reason is the variation in the fitted location of the density peaks, which is transmitted into variation of the peaks of the PUI production rate and count rate. The smallest differences are found for the years during low solar activity. Furthermore, for Ne and O the deviations from the expected longitude seem to follow the decrease of solar activity after 2002, but the increase of activity after the minimum around 2008 is not clearly reflected. This may be a reflection of the weaker upward trend in the ionization rate after 2008 than the downward trend before 2008. In opposite to Ne and O, the peak locations for the He PUI count rate crescent do not present a clear trend in time. Fig.~\ref{figCntRateHeW} shows that it is difficult to unambiguously identify the peak in the crescent region. 

Time variations of the ionization rate are reflected in time variations of the PUI production rate and count rate observed by a detector travelling with Earth. Systematic departures of the fitted peak are due to the general trend in the density series related to the systematic change in the ionization rate. As a result, the structures are not symmetric around the peaks, and this is reflected in the locations of the fitted positions of the peak of the symmetric Gaussian function. Additionally, the deviations present in the results of Gaussian fits to the ISN density and the local PUI production rate for 2011, as well as in the 2008 series of He PUI count rate, arise because of a underdeveloped structure of the crescent which is far from the Gaussian shape and thus the fit is the poorest. The crescent structure is seldom close to the Gaussian shape, it is usually neither symmetric nor smooth in the peak.

\begin{figure}
	\begin{center}
	\includegraphics[width=\columnwidth]{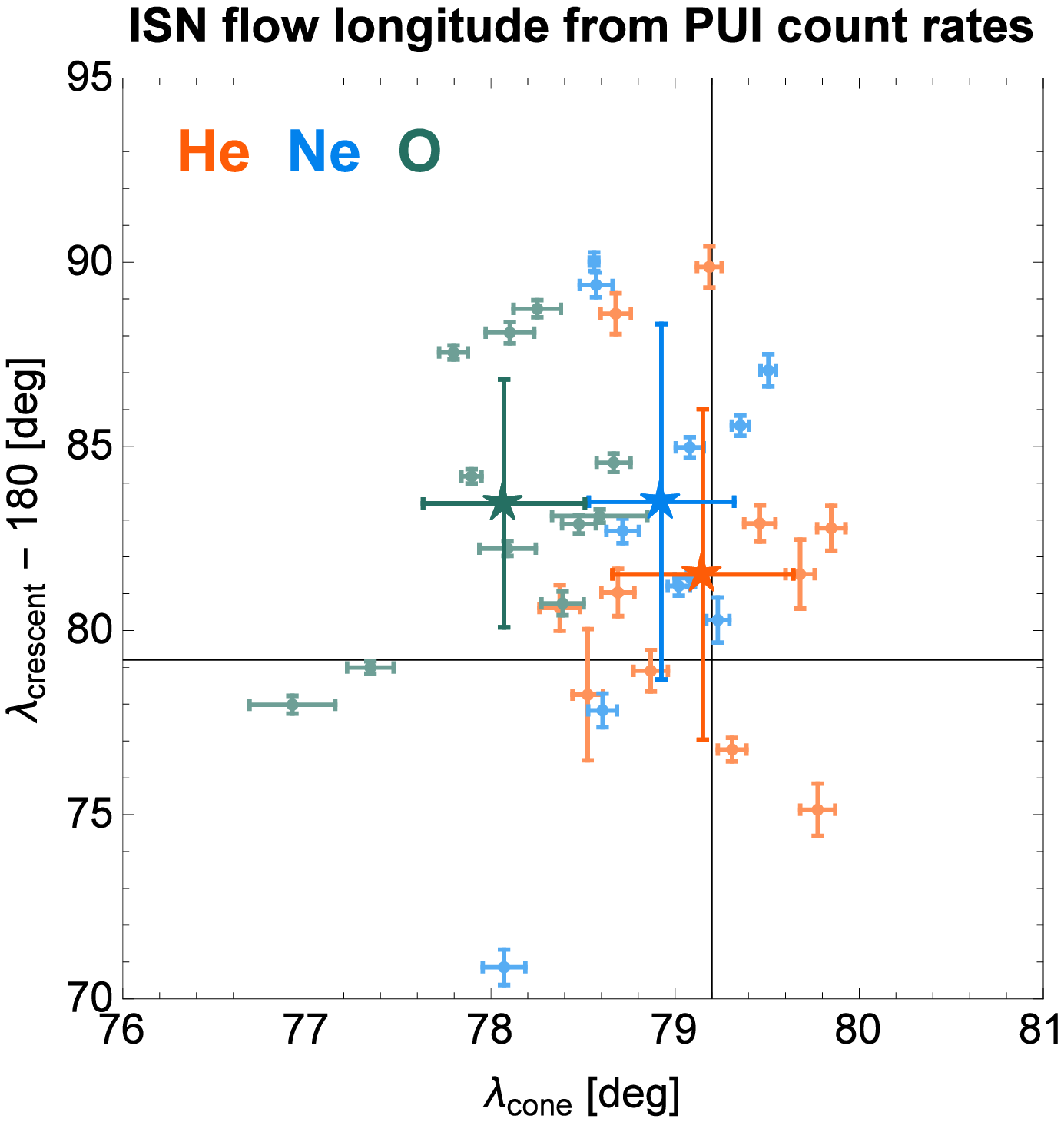}
	\end{center}
	\caption{Longitude ($\lambda$) of the ISN flow derived from the Gaussian fit to the He, Ne, and O PUI count rate cones and crescents for each studied year, together with uncertainties of the fit illustrated as error bars (2013 is omitted because we have results only for crescent as our data set ends before cone creation). The star-points with error bars represent the weighted average value for a given species with weighted standard deviation as an error bar. The cross-hairs mark the longitude expected to reproduce.}
	\label{figISNfittedLong}
\end{figure}

Fig.~\ref{figISNfittedLong} summarizes results of fitting the peak positions for the cone and crescent in the count rate. The points (with respective error bars) correspond to the cone (horizontal axis) and crescent (vertical axis) for a given calendar year. The points with asterisks are the weighted average values of the sample, with the error bars corresponding to the weighted standard deviation of the sample. The averaged cone position for He matches the assumed longitude of the downwind direction almost perfectly. The average longitude for Ne agrees with that assumed within the uncertainty, but the cone position found for O is systematically shifted towards lower longitudes, i.e. oppositely to the difference between the inflow longitude found by \citet{drews_etal:12a} and the longitudes from direct sampling \citep{mccomas_etal:15b}. The peak positions for the crescent are systematically shifted towards larger longitudes for all three species, i.e., in agreement with the direction of the shift resulting from the analysis by \citet{drews_etal:12a}. The magnitudes of the differences are marginally lower than the uncertainties for Ne and He, but not for O. The reason for the systematic differences between the shifts averaged over almost entire solar cycle and the assumed inflow direction are the variations in the ionization rates and (for O) the anisotropies of the solar wind, which are not periodical and cause unbalanced trends, skewing the yearly structure of PUIs enhancement by the Earth.
	% what shifts the peak?

\section{Summary and conclusions}
\label{sec:summary}
\label{sec:conclusions}
We have analysed the modulation of the densities, instantaneous PUI production rates, and simulated PUI count rates of ISN He, Ne, and O for Earth's locations in the ecliptic plane during almost entire solar cycle, from the solar minimum in 2002 until the next maximum in 2013. The density was calculated using the full time- and heliolatitude-dependent hot model of ISN gas inside the heliosphere. The PUI count rate was calculated using the simulated ISN gas as the source population and the classical theory of the creation and evolution of the PUI distribution function in the solar wind, i.e., assuming a distance-independent solar wind speed, instantaneous pitch angle isotropization, and adiabatic cooling. However, instead of the typically made assumption of a zero injection speed of PUIs relative to the Sun, we assumed that their injection speed is equal to the local radial component of the ISN gas bulk velocity, characteristic for the heliocentric distance and ecliptic longitude of the injection location. In addition, we adopted a realistic model of the ISN gas ionization rate, which is based on observations and takes into account all ionization reactions relevant for the three species considered. 

In principle, two maxima in the time series of the PUI count rate are expected during the year: one upwind (crescent), and the other one downwind (cone). The longitudes of the cone and crescent peaks have been used as indicators of the ecliptic longitude of the direction of the ISN gas flow in front of the heliosphere. The goal of our study was to investigate the modulation of the ISN gas density and PUI count rate observed along the Earth's orbit due to the variations of the ionization rates inside heliosphere. Additionally, we wanted to check if the ionization processes could shift the perceived longitude of the inflow direction of ISN gas. 

We simulated the time series of the ISN densities and PUI production and count rates, starting from the classical assumption of spherically symmetric and constant in time ionization rate, and including the assumption of non-zero PUI injection speed. Next, one by one, we switched on the time and heliolatitude variations in the ionization rate and departures of the Earth's orbit from circularity. We found that the crescent can be expected in the PUIs for Ne and O regardless of the details of the production and detection process, but for He, the crescent is created only when the non-zero PUI injection speed is taken into account. In the absence of time- and heliolatitude dependence of the ionization rate, the crescent and cone peaks in the ISN density and PUI production and count rates are found precisely at the upwind and downwind longitudes, respectively. However, switching on the solar cycle modulation of the ionization rate results in spatio-temporal gradients in the ISN gas distribution encountered by Earth in its travel around the Sun, and as a result, some departures of the observed peak positions from the nominal values appear. This is especially true for the PUIs production and count rates. The departures are relatively mild in the cone (almost none for the density and of the order of $2\degr$ for the PUIs) and are much greater for the crescent. 

The positions of the crescent for Ne and O feature a systematic shift towards larger longitudes in 2002, which gradually fades and starts to reverse after $\sim 2008$. The shift in the crescent longitude is appreciably modulated by time variations in the ionization rate. The heliolatitudinal dependence of the ionization rate, which is important mostly for O, adds another systematic effect: the positions of the cone are systematically shifted from their nominal positions. The apparent cone position is shifted towards lower longitudes by $\sim0.5\degr$ in the ISN O density, and in the O PUI production and count rates this shift is more than twice larger. We found that different magnitudes of shifts are because of a secular evolution of the ionization rates. Since during most of this interval the total ionization rates were decreasing (this trend reversed after 2008, and the positive gradient was weaker than the previous negative gradient) and since the solar wind was on average asymmetric in heliolatitude, the departures of the respective longitudes of the crescent and cone from the assumed upwind and downwind longitudes are not zero when averaged over the whole studied time series. The shifts of the crescent are mostly due to the non-balanced temporal trends in the global ionization rate during the calculation interval, and the shift in the cone is due to the departure of the charge exchange and electron-impact ionization rate from spherical symmetry. 

In an ideal case, one could expect that the average of the cone and crescent peak positions over the whole solar cycle should give the expected value. However, it is not so in reality, because of the secular changes in the ionization rates and asymmetries in the solar cycle. In  our case, when averaged over almost the entire solar cycle from 2002 to 2013, the departure of the fitted longitude averaged over the whole sample from the expected value for He is small. Additionally, it is close to the expected value within its respective weighted standard deviation. This is also true for Ne, but with a larger departure. In the case of O, the departure from the expected value is the largest, larger than the standard deviation, which suggests a statistically significant systematic effect. The directions of the departures agree among the species: they are towards larger longitudes for the crescent, in agreement with the direction of the difference found by \citet{drews_etal:12a}, and towards lower longitudes for the cone, oppositely to the difference found by \citet{drews_etal:12a}. Note, however, that in our study the latter ones are within the error bar for He and Ne.

Since our simulations were done for the Earth, while the observations by \citet{drews_etal:12a} were carried out from STEREO, which is on an orbit different from that of the Earth, we cannot say with certainty that it is the solar cycle-related modulation of the ISN gas density and PUI production that is responsible for the difference in the ISN gas inflow direction found by \citet{drews_etal:12a}, but we believe this cause is very likely. The main reason for the effect observed in PUIs is the modulation of the parent ISN gas distribution along the Earth orbit. The short-term variations in the PUI production rates are responsible for additional yearly scatter and tend to increase the observed differences. 

The results we have obtained are particularly interesting in the case of O. They suggest that the heliolatitudinal anisotropy of the ionization rate, resulting from the anisotropy of the solar wind, produces noticeable effects in the distribution of ISN O density in the ecliptic plane and thus must be appropriately taken into account. Thus, the classical model of ISN gas in the heliosphere with a spherically symmetric ionization rate is not sufficient for interpretation of measurements of particle populations related to ISN O.

\appendix
\section{Derivation of the formulae for the PUI flux and count rates}
\label{sec:appendix}
In this appendix, we start from the definition of the PUI flux given in Equation~\ref{eq:puiFluxGeneral} and the relation between the PUI speed at the location of creation $r$ and its speed at a distance of detection $R$ given in Equation~\ref{eq:vPUIsc}, and we calculate the flux and count rate of the PUIs observed at $\Rd$. Equation~\ref{eq:vPUIsc} can be rewritten to the form of $\wSC$ by dividing it by $\vSW\left(R\right)$ on both sides:
	\begin{equation}
	\wSC\left(R\right) = \wSW\left(R\right) \left(\frac{R}{r}\right)^{-\gamma}+1.
	\label{eq:wSC}
	\end{equation}
The flux of PUIs is given by Equation~\ref{eq:puiFluxGeneral}, but in our case the integration must be carried out in the $\wSC$-domain of the ``measured'' PUIs. To do this, $r$ is expressed as a function of $\wSC$ based on Equation~\ref{eq:wSC}:
	\begin{equation}
	r=R\left( \frac{\wSC-1}{\wSW} \right)^{\alpha},
	\label{eq:r-wSC}
	\end{equation}
with $\alpha=\frac{1}{\gamma}$. Finally, with $\mathrm{d}r=\alpha R \wSW^{-\alpha}\left(\wSC-1 \right)^{\alpha-1}\mathrm{d}\wSC$,
the integral from Equation~\ref{eq:puiFluxGeneral} after the change of variables from $r$ to $\wSC$ takes the form:
	\begin{equation}
	F=\alpha R \int\limits_{w_{\mathrm{sc,1}}}^{w_{\mathrm{sc,2}}}{S\wSW^{-3\alpha}\left( \wSC-1 \right)^{3\alpha-1}\mathrm{d}\wSC }.
	\label{eq:puiFlux1}
	\end{equation}
With the normalized PUI speed in the solar wind $\wSW = \vPUIsw / \vSW$, where $\vPUIsw$ is given in Equation~\ref{eq:vPUIsw} and with the PUI source function from Equation~\ref{eq:ProdRate}, we obtain:
	\begin{equation}
	F=\alpha R \int\limits_{w_{\mathrm{sc,1}}}^{w_{\mathrm{sc,2}}}{n \beta\left( 1-\frac{\vrad}{\vSW} \right)^{-3\alpha}\left(\wSC-1\right)^{3\alpha-1}\mathrm{d}\wSC }.
	\label{eq:puiFlux2abs}
	\end{equation}
	
Generally, the differential particle flux takes the form:
	\begin{equation}
	\mathrm{d}F=vf(v)v^2\mathrm{d}v\mathrm{d}\Omega,
	\label{eq:dF}
	\end{equation}
where $f(v)$ is the distribution function of PUIs. The kinetic energy of the measured particle is $E=\frac{1}{2}mv^2$, and $\frac{\mathrm{d}E}{\mathrm{d}v}=mv$, thus:
	\begin{equation}
	\frac{\mathrm{d}F}{\mathrm{d}E\mathrm{d}\Omega}= \frac{v f(v) v^2 \mathrm{d}v \mathrm{d}\Omega}{\mathrm{d}E \mathrm{d}\Omega} = \frac{v f(v) v^2}{m v} = \frac{v^2}{m}f(v).
	\label{eq:dFdE}
	\end{equation}
	
The relation between the count rate $C_p$ and the differential particle flux $\mathrm{d}F$ over the solid angle element $\mathrm{d}\Omega$ and energy range $\mathrm{d}E$ is the following:
	\begin{equation}
	\frac{\mathrm{d}F}{\mathrm{d}E\mathrm{d}\Omega}=\frac{C_p}{\Delta E \, G}
	\label{eq:dF-Cp}
	\end{equation} 
where $\Delta E$ is the energy band width and $G$ a constant effective geometric factor of the instrument. For an electrostatic analyzer $\frac{\Delta (E/Q)}{(E/Q)}$ is constant for a given instrument and determined by its geometry because the measurements are carried out at fixed $E/Q$ steps, where $Q$ is the ionic charge number of the ion. Thus, the energy pass band is determined by
	\begin{equation}
	\Delta E=\frac{\Delta (E/Q)}{(E/Q)}(E/Q)Q,
	\label{eq:DeltaE}
	\end{equation}
Combining Equation~\ref{eq:dF-Cp} and Equation~\ref{eq:DeltaE}, we obtain
	\begin{equation}
	C_p=\frac{\mathrm{d}F}{\mathrm{d}E\mathrm{d}\Omega}E \frac{\Delta (E/Q)}{(E/Q)} G.
	\label{eq:dF-Cp_2}
	\end{equation} 
The quantity $\frac{\mathrm{d}F}{\mathrm{d}E\mathrm{d}\Omega}E$ is the differential energy flux density, and $\frac{\Delta \left( E/Q\right)}{E/Q}G=\textrm{const}=A\,\left[\textrm{m}^2\textrm{sr}\right]$. Thus,
	\begin{equation}
	C_p=\frac{\mathrm{d}F}{\mathrm{d}E\mathrm{d}\Omega}E A.
	\label{eq:dF-Cp_3}
	\end{equation}
Using Equation~\ref{eq:dFdE} we obtain
	\begin{equation}
	C_p= \frac{v^2}{m}f(v) \frac{m}{2} v^2 A = \frac{1}{2} A v^4 f(v) 
	\label{eq:dF-Cp_4}
	\end{equation}
and after applying Equation~\ref{eq:dF} we obtain:	
	\begin{equation}
	C_p=\frac{1}{2} A v \frac{\mathrm{d}F}{\mathrm{d}v \mathrm{d}\Omega}. 
	\label{eq:dF-Cp_5}
	\end{equation}
Now, we transform to the $\wSC$-domain:
	\begin{equation}
	C_p= \frac{1}{2} A \wSC \frac{\mathrm{d}F}{\mathrm{d}\wSC \mathrm{d}\Omega}. 
	\label{eq:dF-Cp_6}
	\end{equation}
From Equation~\ref{eq:dF-Cp_6} it follows that the PUI count rate scales as the differential flux multiplied by the normalized speed of the PUIs. Thus, Equation~\ref{eq:puiFlux2} is modified as follows:
	\begin{equation}
	C_p = \frac{1}{2}\alpha R A \int\limits_{w_{\mathrm{sc,1}}}^{w_{\mathrm{sc,2}}}S\,\wSW^{-3\alpha} \left(\wSC-1\right)^{3\alpha-1}\wSC \mathrm{d}\wSC.
	\label{eq:puiCntRate_Appendix}
	\end{equation}
This form is identical to Equation~\ref{eq:puiCntRate}. $A$ is a proportionality factor and since we are interested in the variation of the count rate along the Earth's orbit and not in its absolute values, we can put it $A=1\left[\textrm{m}^2 \textrm{sr}\right]$. 
	
\section{Calculation scheme}
\label{sec:appendixPart2}
The starting point of the simulations are the densities $n(r,t)$ of the ISN gas, calculated for each day $t$ from 2002 February 9 until 2013 July 20 for distances $r$ from $0.4\Rd$ to $1.1\Rd$ with a step $0.1\Rd$, where $\Rd (t)$ is the Earth's distance from the Sun for the time $t$. The simulations are done as the observer moves with the Earth around the Sun. For each time $t$ the Earth's longitude is calculated and assigned to the computed density. This makes the simulation grid for the densities, which is elliptical in distance and longitude. Since we calculate the densities for each day of the year, the step in longitude varies slightly around the Sun. The densities are calculated with the ionization rates linearly interpolated for each day from the CR-averaged series (dark lines in the left-hand panel of Fig.~\ref{figIonRatesBetaTot}). In calculation of PUI production and count rates we assumed that the ISN density is locally homogeneous in longitude for a given time $t$.

For each day $t$ and distance $r$, the speed of the PUI created by ionization of an ISN atom ($v_{\mathrm{PUI}}^{\mathrm{sw}}$) is calculated using Equation~\ref{eq:vPUIsw}. The radial component of the bulk velocity of ISN neutrals is calculated numerically from the formula $\langle v \rangle_r = \int{v_r f(\textbf{\textit{v}})v^2\mathrm{d}v\mathrm{d}\Omega}$. To calculate $v_{\mathrm{PUI}}^{\mathrm{sw}}$, we use the solar wind speed $v_{\mathrm{sw}}$ for the time $t$ and detection distance $\Rd$, smoothed by the 27 d moving average. With $v_{\mathrm{PUI}}^{\mathrm{sw}}$ and $v_{\mathrm{sw}}$, the speed of PUIs in the spacecraft frame $v_{\mathrm{PUI}}^{\mathrm{sc}}$ is determined following Equation~\ref{eq:vPUIsc}, where the adiabatic cooling of PUIs is taken into account. The ionization rate series used to produce the PUIs is a daily time series for a given time $t$ at 1~au, adjusted to the given distance $r$. The adjustment is because most of the data used to estimate the ionization rates are obtained from spacecraft close to the Earth and no direct information from closer distances to the Sun is available as a homogeneous and long time series. To calculate the ionization conditions at the distance of PUI creation, which is smaller than 1~au, an appropriate adjustment of the series known at 1~au is required in time and space. We assume that PUIs propagate at the solar wind speed typical for the moment of their detection (it is a simplification, but we have to do it because we know only the solar wind speed and density registered at the same moment that PUIs were registered by the detector in the Earth's orbit). For example, a PUI created at a distance $r_j$ is registered at a distance $r_i>r_j$ at a time $t_i$. This allows us to calculate the time span between the creation and the detection under the assumption of a motion at a constant speed $v_i$, it is $\Delta t = (r_j-r_i)/v_i$. Knowing this, we calculate the time $t_j$ of the PUI creation at $r_j$ as $t_j=t_i - \Delta t$. For the found moment $t_j$ we calculate the ionization rate and ISN density and next create the PUI. Additionally, we adjust the EUV flux and solar wind density to account for their quadratic decrease with the distance from the Sun, e.g., solar wind density measured at $r_i$ has to be multiplied by a factor $\left(\frac{r_i}{r_j}\right)^{2}$ to get its value at $r_j$. We are aware that such a simplification of the creation of PUIs is not fully realistic, but without an appropriate MHD modelling it is not possible to improve it. We believe it is sufficient for the study we present given all the other simplifications we make.

The densities ($n$), ionization rates for PUI production ($\beta$), and radial speeds of the ISN atoms ($\vrad$) are tabulated as a function of distance $r$ and time $t$. In the calculation of count rates based on Equation~\ref{eq:puiCntRate}, we tabulate the integrand function as a function of $\wSC$ and adopt a linear interpolation to the initial grid to determine the count rates for the ions created at a given distance $r$ and time $t$. Finally, the count rates are normalized by an average value over the whole time series studied for each species separately.

\section*{Acknowledgements}
The authors from SRC PAS acknowledge the support by the grant 2012/06/M/ST9/00455 from the National Science Center, Poland. The OMNI data were obtained from the GSFC/SPDF OMNIWeb interface at http://omniweb.gsfc.nasa.gov. We acknowledge LASP for the release of TIMED/SEE data for public use and ISEE (Nagoya University, Japan) for the solar wind speed data from the interplanetary scintillation observations.

\bibliographystyle{mnras}
\bibliography{iplbib}{}

% Don't change these lines
\bsp	% typesetting comment
\label{lastpage}
\end{document}